\documentclass[authoryear,review,12pt]{elsarticle}

\usepackage{amssymb}
\usepackage{float}
\usepackage{booktabs,tabularx}
\usepackage{amsmath}
\usepackage[utf8]{inputenc}
\usepackage[english]{babel}
\usepackage[letterpaper,top=2cm,bottom=2cm,left=3cm,right=3cm]{geometry}
\usepackage{graphicx}
\usepackage{cancel}
\usepackage{bm}
\usepackage{enumitem}
\usepackage[colorlinks=true, allcolors=blue]{hyperref}
\usepackage{comment}
\usepackage{newunicodechar}
\usepackage{caption}
\usepackage{array}
\usepackage[table]{xcolor}
\usepackage{makecell}
\usepackage{tikz}
\usetikzlibrary{arrows.meta,positioning,calc,fit,backgrounds,decorations.pathmorphing}
\newcommand{\hspring}[3]{%
   \draw[thick] ({#1},{#3}) -- ({#1+0.15},{#3});
   \draw[thick,decorate,decoration={zigzag,segment length=2.8pt,amplitude=3.2pt}]
        ({#1+0.15},{#3}) -- ({#2-0.15},{#3});
   \draw[thick] ({#2-0.15},{#3}) -- ({#2},{#3});}
\newcommand{\hdashpot}[3]{%
   \pgfmathsetmacro{\m}{(#1+#2)/2}
   \draw[thick] ({#1},{#3}) -- ({\m-0.2},{#3});
   \draw[thick] ({\m-0.2},{#3+0.2}) -- ({\m-0.2},{#3-0.2}) -- ({\m+0.2},{#3-0.2});
   \draw[thick] ({\m-0.2},{#3+0.2}) -- ({\m+0.2},{#3+0.2});
   \draw[thick] ({\m+0.05},{#3+0.14}) -- ({\m+0.05},{#3-0.14});
   \draw[thick] ({\m+0.05},{#3}) -- ({#2},{#3});}

\journal{Journal of the Mechanics and Physics of Solids}

\begin{document}

\begin{frontmatter}

\title{Multiphysical impedance spectroscopy of porous electrodes based on linear irreversible thermodynamics}

\author{Junning Jiao}
\ead{jiao.ju@northeastern.edu}

\author{Juner Zhu\corref{cor1}}
\ead{j.zhu@northeastern.edu}
\cortext[cor1]{Corresponding author.}

\affiliation{organization={Department of Mechanical and Industrial Engineering, Northeastern University},
            addressline={360 Huntington Ave},
            city={Boston},
            postcode={02115},
            state={MA},
            country={USA}}

\begin{abstract}
Porous electrodes couple electrical, chemical, mechanical, hydraulic, and
thermal fields, yet conventional frequency-domain diagnostics interrogate
only one of them: electrochemical impedance spectroscopy (EIS) the
electrical response and dynamic mechanical analysis (DMA) the mechanical.
Each reads a diagonal entry of the multiphysical constitutive matrix and
is blind to the cross-couplings that govern structural evolution and
degradation. Starting from linear irreversible thermodynamics, we
formulate a general theory of multiphysical impedance spectroscopy, in
which perturbing one field and measuring the conjugate response of
another probes an off-diagonal entry of the constitutive (Onsager)
matrix, recovering the static coupling coefficient and resolving its
relaxation dynamics across frequency. Specializing to the
electro-chemo-mechanical pathway yields a closed-form theory of
mechano-electrochemical impedance spectroscopy (MEIS), in which a small
harmonic current is applied and the stack stress is measured; the
impedance factorizes into a chemical-accumulation term multiplying the
sum of a chemo-mechanical and a poro-mechanical kernel. The
porosity-accommodation bridge function is derived from a Helmholtz free
energy---following from a microstructural stiffness and viscosity rather
than a fitted form---and a three-phase (solid--fluid--void) closure
interpolates continuously between unsaturated and Biot-saturated limits
through a void-accommodation fraction. Non-dimensionalization reduces the
spectrum to five groups, identifies the phase angle as the discriminator
of the chemo-mechanical parameters, and locates the onset of
second-quadrant behavior, which in a full cell arises from the
competition between an expanding and a contracting electrode. Because the
mechanical kernel reduces to an equivalent spring--dashpot network,
measured spectra can be fitted as in EIS or DMA with each element
retaining a definite physical meaning, and MEIS emerges as one member of
a family of cross-coupled spectroscopies the same framework brings within
reach.
\end{abstract}


\end{frontmatter}

\section{Introduction}
\label{sec1}

Many natural and engineered systems are governed by the simultaneous 
action of multiple physical fields. Electrical, chemical, mechanical, 
hydraulic, and thermal processes rarely act in isolation; instead, they 
couple, so that a gradient in one field drives a flux in another~\citep{prigogine_introduction_1963}. 
Table~\ref{tab:cross_effects} qualitatively summarizes various direct and cross-coupling effects among electric, chemical, mechanical (solids and fluids), and thermal fields. 
For example, thermoelectricity converts heat flow into electric current; 
electro-osmosis drives fluid through a charged porous medium under an 
electric field; thermal expansion couples temperature to deformation; 
and diffusion-induced stress couples composition to mechanics. Such 
cross-couplings are not incidental complications but are central to the 
function of devices ranging from fuel cells and thermoelectric 
generators to actuators, sensors, and biological tissues. The 
theoretical description of these couplings is the domain of linear 
irreversible thermodynamics, which, through the Onsager reciprocal 
relations~\citep{onsager_reciprocal_1931,onsager_reciprocal_1931II}, 
provides a unified and thermodynamically consistent structure relating 
the full set of generalized fluxes to their conjugate driving forces~\citep{balluffi2005kinetics}.

\definecolor{diagblue}{RGB}{222,235,247}
\definecolor{headgray}{RGB}{242,242,242}

\newcolumntype{C}[1]{>{\centering\arraybackslash}m{#1}}
\newcolumntype{L}[1]{>{\raggedright\arraybackslash\bfseries}m{#1}}

\begin{table*}[t]
  \centering
  \footnotesize
  \renewcommand\arraystretch{1.45}
  \renewcommand\theadfont{\bfseries\footnotesize}
  \setlength\tabcolsep{4pt}
  \begin{tabular}{|L{2.2cm}|*{5}{C{2.45cm}|}}
    \hline
    \rowcolor{headgray}
    \thead{Potential\\$\backslash$\,Flux} &
    \thead{Electric current\\$\bm{J}_e$} &
    \thead{Ionic flux\\$\bm{F}_i$} &
    \thead{Deformation\\$\dot{\bm{\varepsilon}}^{p}$} &
    \thead{Fluid flow\\$\bm{v}_f$} &
    \thead{Heat flux\\$\bm{J}_q$}\\
    \hline
    Electric potential $\phi$
     & \cellcolor{diagblue}Electronic conduction & Electromigration & Electrostriction & Electro-osmosis & Peltier effect\\
    \hline
    Chemical potential $\mu$
     & Diffusion potential & \cellcolor{diagblue}Ionic diffusion & Compositional expansion & Chemo-osmosis & Dufour effect\\
    \hline
    Stress $\bm{\sigma}^{p}$
     & Piezoelectricity & Stress-assisted diffusion & \cellcolor{diagblue}Skeleton viscoelasticity & Poroelastic consolidation & Piezocaloric effect\\
    \hline
    Fluid pressure $p$
     & Streaming current & Advective ion transport & Pore-pressure swelling & \cellcolor{diagblue}Darcy flow & Thermal advection\\
    \hline
    Temperature $T$
     & Seebeck effect & Soret effect & Thermal expansion & Thermo-osmosis & \cellcolor{diagblue}Heat conduction\\
    \hline
  \end{tabular}
  \caption{Direct (shaded, on the diagonal) and cross-coupling effects among the five
    coupled processes in a porous electrode. Each row is a thermodynamic driving force
    (potential gradient) and each column the conjugate generalized flux; the entry names
    the physical effect by which a given potential drives a given flux. The constitutive
    relations underlying these effects are developed in the following sections.}
  \label{tab:cross_effects}
\end{table*}

Among multiphysical systems, porous electrodes in electrochemical 
energy technologies are one of the most strongly coupled. They are 
central to lithium-ion batteries, electrochemical capacitors, fuel 
cells, electrolysis systems, and direct lithium 
extraction~\citep{newman_porous-electrode_1975,lai_3d_2018,
huang_reviewimpedance_2020,chen_porous_2022}, and their operation 
simultaneously involves all five of the physical fields named above. 
During operation, electrons are conducted through the solid matrix; 
ions migrate and diffuse through the electrolyte-filled pore network; 
electrochemical reactions at solid--electrolyte interfaces drive 
compositional changes; these compositional changes deform the solid 
skeleton; the deformation in turn drives flow of the pore fluid; and 
the entire process generates and is modulated by heat. A porous 
electrode is therefore a system in which electrical, chemical, 
mechanical, hydraulic, and thermal fields are intrinsically and 
bidirectionally coupled within a single heterogeneous microstructure 
composed of solid active materials, electrolyte-filled pores, and 
conductive additives.

These couplings are not merely of academic interest; they govern 
performance and degradation. For example, intercalation-induced 
swelling of active particles produces stress accumulation, particle 
fracture, electrode-level deformation, and progressive degradation of 
the porous structure~\citep{ipers_rapid_2024,jiao_modeling_2026}. 
Reciprocally, changes in porosity and pore morphology alter electrolyte 
transport and reaction kinetics, feeding back on the electrochemical 
response~\citep{xu_guiding_2021,song_microstructural_2025, zhao_electrolytes_nodate}. Capturing this web of interactions requires a framework that 
treats all the participating fields and their couplings on an equal 
footing.

\paragraph{Modeling of porous electrodes}
On the electrochemical side, continuum descriptions of porous 
electrodes have been largely shaped by the pioneering work of Newman 
and collaborators, now known as porous electrode theory 
(PET)~\citep{newman_porous-electrode_1975,doyle_modeling_1993,
fuller_electrochemical_2018}. PET provides a reduced-order description 
of electrochemical processes by averaging microscale transport and 
reaction phenomena over the porous structure, and has become the 
foundation for modeling a wide variety of electrochemical systems. 
Subsequent developments incorporated concentrated-solution effects, 
phase transitions, and nonlinear reaction kinetics; in particular, 
nonequilibrium thermodynamic formulations have captured phase 
separation, electro-autocatalysis, and mosaic instabilities in 
battery materials~\citep{bazant_theory_2013,
Bazant2017ThermodynamicElectro-autocatalysis,Smith2017MultiphaseTheory,
lian_modeling_2024}. In parallel, the mechanics community has developed 
thermodynamically consistent theories of chemo-mechanical coupling in 
solids~\citep{di_leo_cahnhilliard-type_2014,chester_finite_2015,
klinsmann_dendritic_2019}, treating diffusion as a thermodynamically 
driven process influenced by mechanical fields, and providing insight 
into stress-assisted diffusion, phase transformations, particle 
fracture, and diffusion-induced stresses. These two lines of work have 
greatly advanced predictive capability, but they have largely developed 
the electrochemical and mechanical descriptions separately, and 
high-fidelity models have grown increasingly costly and parameter-rich, 
making direct identification of physical parameters from low-cost 
experiments difficult.

\paragraph{Diagnostics and the impedance concept}
A complementary route to understanding coupled systems is provided by 
frequency-domain diagnostics. The most established is electrochemical 
impedance spectroscopy (EIS)~\citep{huang_reviewimpedance_2020,
wang_electrochemical_2021}, in which a small harmonic electrical 
perturbation is applied and the electrical response is measured, 
yielding a transfer function whose frequency dependence separates 
distinct transport and kinetic processes. The mechanical analogue is 
dynamic mechanical analysis (DMA)~\citep{menard2020dynamic,
ferry1961viscoelastic}, in which a small harmonic strain is imposed 
and the resulting stress is recorded, characterizing viscoelastic 
relaxation. Both techniques share a common principle: a small periodic 
perturbation in one field, a measured periodic response in a conjugate 
field, and a frequency-domain transfer function that resolves the 
underlying dynamics. EIS and DMA, however, each remain confined to a 
single physical domain---EIS to the electrical, DMA to the 
mechanical---and therefore probe only the diagonal, self-coupled 
response of the system. Neither directly measures the cross-coupling 
between different physical fields, even though it is precisely these cross-couplings that 
encode the multiphysical interactions of interest.

\paragraph{A unifying view: multiphysical impedance spectroscopy}
Viewed through the lens of linear irreversible thermodynamics, EIS and 
DMA are revealed as two members of a much larger family. The multiphysical matrix in Table~\ref{tab:cross_effects} relates every generalized flux to every conjugate 
force; its diagonal entries correspond to the classical transport laws 
(Ohm, Fick, Newton, Darcy, Fourier, see Eq.~\ref{eq:consmatrix} of the following section for details), while its off-diagonal entries 
correspond to the multiphysical cross-couplings. Each entry of this matrix, in 
principle, defines a dynamic spectroscopy: one drives a generalized 
flux harmonically and measures a conjugate response, reading the 
transfer function from the corresponding matrix structure. EIS probes 
the electrical diagonal; DMA probes the mechanical diagonal. The 
\emph{off-diagonal} entries define a class of \emph{cross-coupled} 
spectroscopies that probe the multiphysical interactions directly, 
and that remain almost entirely unexplored. We refer to this general 
class as \emph{multiphysical impedance spectroscopy}.

Recently, our group introduced one realization of this class: 
mechano-electrochemical impedance spectroscopy 
(MEIS)~\citep{fang_mechano-electrochemical_2025}, in which a small 
harmonic electric current is applied and the mechanical (pressure or 
stress) response is measured. MEIS occupies an off-diagonal pathway of 
the multiphysical constitutive matrix, coupling the electrical, chemical, and mechanical 
fields, and thereby provides access to structural evolution---porosity 
change, particle rearrangement, electrode consolidation, stress 
accumulation---that is invisible to purely electrical measurements. As 
a low-cost, noninvasive probe that requires only the addition of a 
force sensor to a standard electrochemical setup, MEIS is practical and 
broadly deployable. Despite a strict derivation of an equivalent phenomenological model of MEIS has been available, 
its theoretical foundation, and more 
fundamentally the theory of the general multiphysical-impedance class 
to which it belongs, has not yet been established.

\paragraph{This work}
In this work, we develop a general framework for \emph{multiphysical 
impedance spectroscopy} of porous electrodes based on the linear theory 
of irreversible thermodynamics. Starting from the full multiphysical 
constitutive matrix coupling electrical, chemical, mechanical, 
hydraulic, and thermal fields, we identify the family of dynamic 
spectroscopies generated by its diagonal and off-diagonal entries, and 
derive the corresponding frequency-domain transfer functions. We then 
specialize the framework to the electro-chemo-mechanical pathway, 
yielding a closed-form theory for MEIS that explicitly accounts for 
chemo-mechanical coupling, viscoelastic relaxation of the solid 
skeleton, and---through a three-phase poromechanical treatment of the 
partially saturated electrode---the coupled evolution of pore-volume 
change and pore-fluid pressurization. The theory provides a 
thermodynamically consistent and physically interpretable foundation 
for MEIS, identifies the characteristic relaxation time scales 
governing the measured spectra, and recovers existing phenomenological 
models as limiting cases. More broadly, it positions MEIS as one 
member of a family of multiphysical impedance spectroscopies and 
points toward additional, as-yet-unrealized cross-coupled measurements 
that the same framework predicts. Although restricted to the linear 
regime, the framework is naturally suited to small-perturbation 
spectroscopies and applies equally to EIS, DMA, and their 
cross-coupled counterparts.

\section{General Multiphysical Framework}
\label{sec: General Framework}

\subsection{Generalized Fluxes and Driving Forces}

\begin{figure}[htbp]
    \centering
    \includegraphics[width=0.8\linewidth]{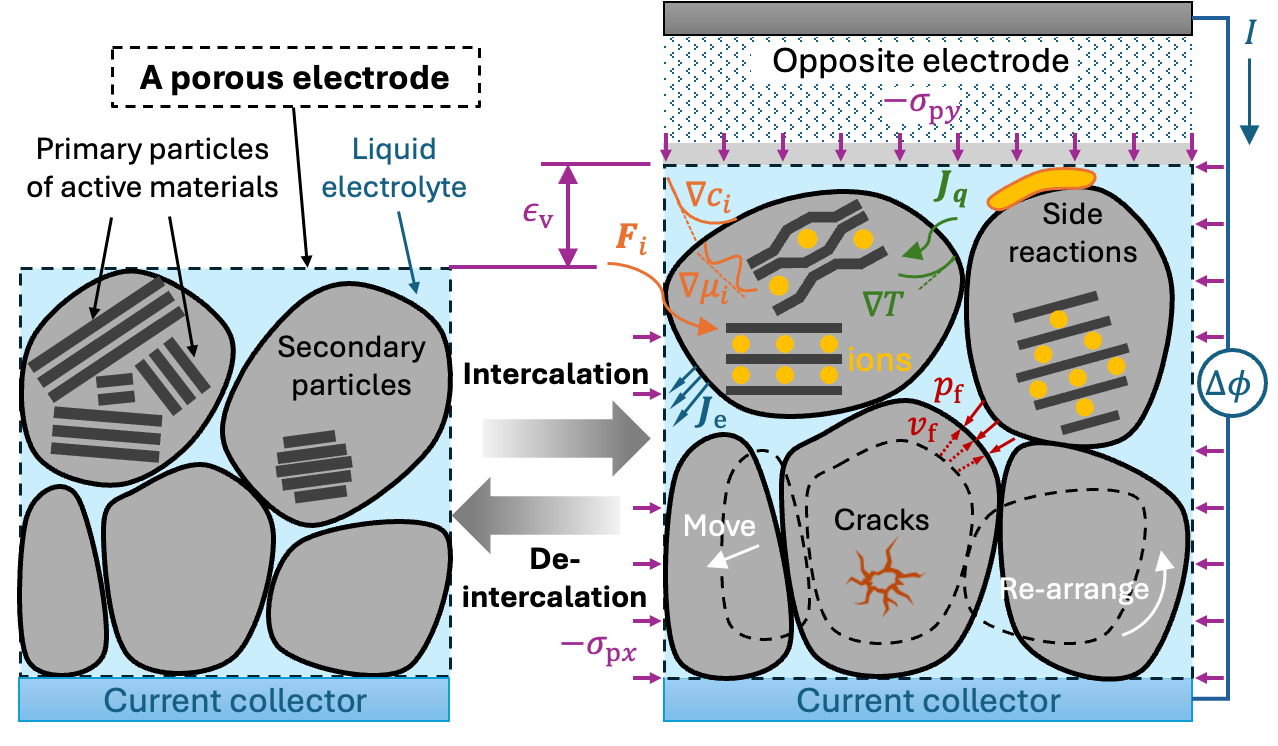}
    \caption{Illustration of the coupled processes in a porous electrode. 
    Fluxes: $\mathbf{J}_e$, electric current density; $\mathbf{F}_i$, 
    ionic flux; $\dot{\bm{\varepsilon}}^p$, strain rate of the porous 
    medium; $\mathbf{v}_f$, fluid velocity; $\mathbf{J}_q$, heat flux. 
    Forces: $\phi$, electric potential; $c_i$, ionic concentration; 
    $\bm{\sigma}^p$, stress of the porous medium; $p$, fluid pressure; 
    $T$, temperature.}
    \label{fig:PorousElectrodes}
\end{figure}

We consider a porous electrode undergoing coupled electrical, chemical, 
mechanical, hydraulic, and thermal processes, as illustrated in 
Figure~\ref{fig:PorousElectrodes}. The active material exists in the 
form of particles. In typical industrial applications, each particle 
is itself an agglomerate of many smaller primary particles and is 
therefore referred to as a secondary particle. These particles form a 
connected network--hereinafter called the solid skeleton--immersed in a 
liquid electrolyte.

The framework considers five irreversible processes: electrical 
conduction, ionic transport, mechanical dissipation, fluid flow, and 
heat transport. Each process is characterized by a generalized flux 
and a conjugate thermodynamic driving force. We assemble the generalized 
fluxes into the vector
\begin{equation}
\mathbf{J} 
=
\begin{bmatrix}
\mathbf{J}_e &
\mathbf{F}_i &
\dot{\bm{\varepsilon}}^{p} &
\mathbf{v}_f &
\mathbf{J}_q
\end{bmatrix}^{\mathrm{T}},
\end{equation}
where $\mathbf{J}_e = (J_{e1}, J_{e2}, J_{e3})$ is the electric current 
density, $\mathbf{F}_i = (F_{i1}, F_{i2}, F_{i3})$ is the ionic flux, 
$\dot{\bm{\varepsilon}}^{p} = (\dot{\varepsilon}^{p}_1, 
\dot{\varepsilon}^{p}_2, \dot{\varepsilon}^{p}_3)$ are the principal 
strain rates of the porous medium, $\mathbf{v}_f = (v_{f1}, v_{f2}, 
v_{f3})$ is the fluid velocity, and $\mathbf{J}_q = (J_{q1}, J_{q2}, 
J_{q3})$ is the heat flux. 

The corresponding conjugate thermodynamic 
driving forces are
\begin{equation}
\mathbf{X} =
\begin{bmatrix}
-\nabla \phi &
-\nabla \mu &
-\bm{\sigma}^{p} &
-\nabla p &
-\nabla T
\end{bmatrix}^{\mathrm{T}},
\end{equation}
where $-\nabla \phi$ is the electric field driving the electric current, 
$-\nabla \mu$ is the chemical potential gradient driving ionic transport, 
$-\bm{\sigma}^{p} = -(\sigma_1^p, \sigma_2^p, \sigma_3^p)$ are the 
principal stresses of the porous medium conjugate to the strain rate, 
$-\nabla p$ is the pressure gradient driving fluid flow, and $-\nabla T$ 
is the temperature gradient driving heat transport. The negative sign in front of $\bm{\sigma}^p$ reflects the standard 
solid-mechanics convention in which stress is positive in tension, so 
that compressive loading---driving inward strain---corresponds to a 
positive driving force in the entropy production, which will be discuss in Section~\ref{sec:entropyprod}. This convention 
ensures that all entries of $\mathbf{X}$ are uniformly signed: each 
component represents the driving force whose product with the conjugate 
flux contributes positively to the entropy production rate.

A remark on the mechanical variables is in order. Rather than working 
with the full stress and strain-rate tensors, we have restricted the 
mechanical entries of $\mathbf{J}$ and $\mathbf{X}$ to their principal 
components. This simplification is justified by two considerations. 
First, we assume that the porous electrode is statistically isotropic 
at the homogenized scale: the active particles and the pore network 
have no preferred orientation, and the constitutive response is 
invariant under rotation. In an isotropic medium, the stress and 
strain-rate tensors are coaxial, so the constitutive relations between 
them reduce to relations between principal values along the common 
principal axes~\citep{coussy2004poromechanics}. Moreover, the 
\textit{Curie--Prigogine principle} of linear irreversible 
thermodynamics states that, in an isotropic system, fluxes and forces 
of different tensorial rank cannot couple~\citep{degroot1984non-equilibrium,
demirel2019nonequilibrium}. Consequently, the scalar driving forces 
($-\nabla\phi$, $-\nabla\mu$, $-\nabla p$, $-\nabla T$) couple only to 
the hydrostatic (scalar) part of the mechanical response, and the 
deviatoric (shear) components are decoupled from the other physical 
fields. Second, in the experimental configurations of primary interest, 
including stack-pressure measurements and 
MEIS~\citep{fang_mechano-electrochemical_2025}, the electrode is 
mechanically constrained along a single direction and the measured 
mechanical quantity is the normal stress (or pressure) along the 
stacking axis. The principal-stress formulation thus captures the 
mechanical degrees of freedom that are both physically relevant for 
multiphysical coupling and experimentally accessible.

\subsection{Assumptions and Linear Response Regime}

The formulation of theoretical framework is developed under the following assumptions, standard 
in linear irreversible thermodynamics~\citep{balluffi2005kinetics}:

\begin{enumerate}
    \item \textbf{Small perturbations about a local reference state.} 
    All state variables deviate only slightly from a thermodynamic 
    reference state, so that the system response remains within the 
    linear regime and superposition applies.
    
    \item \textbf{Local thermodynamic equilibrium.} Although the system 
    is globally out of equilibrium, each material point is in local 
    equilibrium, so that thermodynamic state variables such as 
    temperature, chemical potential, and stress remain well defined 
    throughout the continuum.
    
    \item \textbf{Small-strain kinematics.} Deformation is described by 
    the infinitesimal strain tensor, and the governing equations are 
    linearized about a reference configuration of uniform temperature. 
    Temperature gradients are retained as thermodynamic driving forces 
    for heat transport.
    
    \item \textbf{Onsager reciprocity.} The microscopic dynamics are 
    assumed to satisfy time-reversal symmetry, so that the matrix of 
    phenomenological coefficients relating fluxes to conjugate forces 
    is symmetric~\citep{onsager_reciprocal_1931,onsager_reciprocal_1931II}.
\end{enumerate}

\subsection{Constitutive Relations}

Under the assumptions stated above, the generalized fluxes depend 
linearly on the thermodynamic driving forces. The constitutive 
relation is therefore written as
\begin{equation}
\mathbf{J} = \mathbb{L}\,\mathbf{X},
\end{equation}
where $\mathbb{L}$ is the constitutive matrix of phenomenological 
coefficients. Explicitly,
\begin{equation}
\begin{bmatrix}
\mathbf{J}_e \\
\mathbf{F}_i \\
\dot{\bm{\varepsilon}}^{p} \\
\mathbf{v}_f \\
\mathbf{J}_q
\end{bmatrix}
=
\begin{bmatrix}
\mathbf{L}_{11} & \mathbf{L}_{12} & \mathbf{L}_{13} & \mathbf{L}_{14} & \mathbf{L}_{15} \\
\mathbf{L}_{21} & \mathbf{L}_{22} & \mathbf{L}_{23} & \mathbf{L}_{24} & \mathbf{L}_{25} \\
\mathbf{L}_{31} & \mathbf{L}_{32} & \mathbf{L}_{33} & \mathbf{L}_{34} & \mathbf{L}_{35} \\
\mathbf{L}_{41} & \mathbf{L}_{42} & \mathbf{L}_{43} & \mathbf{L}_{44} & \mathbf{L}_{45} \\
\mathbf{L}_{51} & \mathbf{L}_{52} & \mathbf{L}_{53} & \mathbf{L}_{54} & \mathbf{L}_{55}
\end{bmatrix}
\begin{bmatrix}
-\nabla \phi \\
-\nabla \mu \\
-\bm{\sigma}^{p} \\
-\nabla p \\
-\nabla T
\end{bmatrix}.
\label{eq:consmatrix}
\end{equation}
Each block $\mathbf{L}_{ij}$ is a second-order tensor mapping a 
three-component driving force to a three-component flux. The diagonal 
blocks $\mathbf{L}_{ii}$ correspond to the classical transport laws 
associated with each physical process---Ohm's law, Fick's law, 
Newton's law of viscosity, Darcy's law, and Fourier's law---while the 
off-diagonal blocks $\mathbf{L}_{ij}$ ($i \neq j$) encode the 
cross-couplings between distinct physical mechanisms. Onsager 
reciprocity further constrains the matrix to be symmetric,
\begin{equation}
\mathbf{L}_{ij} = \mathbf{L}_{ji}^{\mathrm{T}},
\end{equation}
reducing the number of independent coupling coefficients. The physical 
content of each block is examined in 
Section~\ref{sec:termsofmatrix}.

\subsection{Thermodynamic Admissibility}
\label{sec:entropyprod}

The constitutive law must satisfy the second law of thermodynamics, 
which requires non-negative local entropy production. With the 
generalized fluxes and conjugate forces introduced above, the local 
entropy production rate takes the bilinear form
\begin{equation}
\dot{s}_{\text{prod}} = \mathbf{J}^{\mathrm{T}} \mathbf{X}.
\label{eq:entropyprod_bilinear}
\end{equation}
Substituting the linear constitutive relation $\mathbf{J} = \mathbb{L}\,\mathbf{X}$ gives
\begin{equation}
\dot{s}_{\text{prod}} = \mathbf{X}^{\mathrm{T}}\,\mathbb{L}\,\mathbf{X} \geq 0.
\label{eq:entropyprod_quadratic}
\end{equation}
For this inequality to hold for all admissible driving forces 
$\mathbf{X}$, the symmetric part of $\mathbb{L}$ must be positive 
semi-definite. Combined with the Onsager reciprocity introduced in the 
preceding subsection, which renders $\mathbb{L}$ symmetric, the 
constitutive matrix itself must therefore be positive semi-definite:
\begin{equation}
\mathbb{L} = \mathbb{L}^{\mathrm{T}}, 
\qquad
\mathbf{X}^{\mathrm{T}}\,\mathbb{L}\,\mathbf{X} \geq 0 
\quad \forall\,\mathbf{X}.
\end{equation}
These two conditions---symmetry and positive semi-definiteness---together 
constitute the thermodynamic admissibility requirements on the 
phenomenological coefficients. They imply that each diagonal block 
$\mathbf{L}_{ii}$ is itself positive semi-definite, ensuring that 
every transport process is individually dissipative, and they bound 
the off-diagonal coupling blocks $\mathbf{L}_{ij}$ in terms of the 
diagonal blocks. For scalar coefficients, this constraint takes the 
familiar form
\begin{equation}
L_{ij}^{2} \leq L_{ii}\,L_{jj},
\label{eq:cauchy_schwarz_onsager}
\end{equation}
which is a direct consequence of positive semi-definiteness and 
limits the magnitude of any cross-coupling by the geometric mean of 
the associated direct dissipation coefficients~\citep{degroot1984non-equilibrium,
demirel2019nonequilibrium}.

\subsection{Classical Transport Laws and Cross-Couplings}
\label{sec:termsofmatrix}

The constitutive matrix $\mathbb{L}$ introduced in the previous 
subsections naturally recovers the classical phenomenological transport 
laws as its diagonal blocks, while the off-diagonal blocks encode the 
multiphysical couplings between distinct transport processes.

\paragraph{Diagonal blocks: classical transport laws}
The diagonal block $\mathbf{L}_{11}$ relates the electric current 
density to the electric field,
\begin{equation}
\mathbf{J}_e = \mathbf{L}_{11}(-\nabla \phi),
\end{equation}
which is Ohm's law with $\mathbf{L}_{11}$ playing the role of the 
electrical conductivity tensor. Similarly, $\mathbf{L}_{22}$ governs 
ionic transport,
\begin{equation}
\mathbf{F}_i = \mathbf{L}_{22}(-\nabla \mu).
\end{equation}

For dilute solutions, $\mu = \mu_0 + RT\ln c$ gives 
$\nabla\mu = (RT/c)\nabla c$, where $c$ is the concentration of the intercalation ions, and the transport law reduces to
\begin{equation}
\mathbf{F}_i = -\mathbf{D}\,\nabla c,
\end{equation}
recovering Fick's law of diffusion with the effective diffusivity 
$\mathbf{D} = \mathbf{L}_{22}\,RT/c$.

The block $\mathbf{L}_{33}$ describes the dissipative mechanical 
response of the porous medium as a whole,
\begin{equation}
\dot{\bm{\varepsilon}}^{p} = \mathbf{L}_{33}(-\bm{\sigma}^{p}),
\end{equation}
which is Newton's law of viscosity, with $\mathbf{L}_{33}$ playing 
the role of an inverse viscosity tensor. The superscript $p$ 
emphasizes that both the strain rate and the stress are defined at 
the homogenized porous-medium scale rather than on the solid skeleton 
alone, and thereby reflect the combined dissipative behavior of the 
skeleton, the pore fluid, and their interactions. Fluid transport through the porous medium is described 
by $\mathbf{L}_{44}$,
\begin{equation}
\mathbf{v}_f = \mathbf{L}_{44}(-\nabla p),
\end{equation}
which is Darcy's law with $\mathbf{L}_{44} = \bm{\kappa}/\eta_f$, where 
$\bm{\kappa}$ is the permeability tensor and $\eta_f$ is the electrolyte 
viscosity. Finally, $\mathbf{L}_{55}$ governs heat conduction,
\begin{equation}
\mathbf{J}_q = \mathbf{L}_{55}(-\nabla T),
\end{equation}
which is Fourier's law with $\mathbf{L}_{55}$ the thermal conductivity 
tensor.

\paragraph{Off-diagonal blocks: cross-couplings}
The off-diagonal blocks $\mathbf{L}_{ij}$ ($i \neq j$) capture the 
multiphysical couplings that distinguish the present framework from 
the uncoupled classical laws. Each pair $\mathbf{L}_{ij} = 
\mathbf{L}_{ji}^{\mathrm{T}}$ encodes a reciprocal cross-effect between 
two distinct processes. Several couplings are of particular relevance 
for porous electrodes:

\begin{itemize}
    \item \textbf{Electro-chemical coupling} ($\mathbf{L}_{12}$, 
    $\mathbf{L}_{21}$): chemical potential gradients contribute to 
    electric current (diffusion potential), while electric fields drive 
    ion migration (electromigration).
    
    \item \textbf{Chemo-mechanical coupling} ($\mathbf{L}_{23}$, 
    $\mathbf{L}_{32}$): chemical potential gradients drive deformation 
    (intercalation-induced expansion), while stresses bias ionic 
    transport (stress-assisted diffusion).
    
    \item \textbf{Poro-mechanical coupling} ($\mathbf{L}_{34}$, 
    $\mathbf{L}_{43}$): pore-fluid pressure gradients deform the solid 
    skeleton, while skeletal deformation drives fluid flow within the 
    pore network.
    
    \item \textbf{Electrokinetic coupling} ($\mathbf{L}_{14}$, 
    $\mathbf{L}_{41}$): electric fields drive fluid flow (electroosmosis) 
    and fluid flow generates electric currents (streaming potential).
    
    \item \textbf{Thermo-diffusive coupling} ($\mathbf{L}_{25}$, 
    $\mathbf{L}_{52}$): temperature gradients drive species transport 
    (Soret effect), while concentration gradients contribute to heat 
    transport (Dufour effect).
\end{itemize}

Additional couplings, such as thermoelectric ($\mathbf{L}_{15}$, 
Seebeck/Peltier effects) and thermomechanical 
($\mathbf{L}_{35}$, thermal expansion) interactions, are likewise 
captured within the framework. Some of these couplings may vanish due 
to material symmetry, the Curie--Prigogine principle, or the specific 
physical structure of the system; the surviving couplings constitute 
the multiphysical signature of the porous electrode.

\subsection{Quadratic Dissipation Potential}

An equivalent and more fundamental formulation of the linear constitutive 
relations is obtained by introducing a scalar dissipation 
potential~\citep{ziegler1983introduction,germain1983continuum}. We 
define
\begin{equation}
\Phi(\mathbf{X}) = \frac{1}{2}\,\mathbf{X}^{\mathrm{T}}\,\mathbb{L}\,\mathbf{X},
\label{eq:dissipation_potential}
\end{equation}
so that the generalized fluxes are recovered by differentiation,
\begin{equation}
\mathbf{J} = \frac{\partial \Phi}{\partial \mathbf{X}} = \mathbb{L}\,\mathbf{X}.
\label{eq:flux_from_potential}
\end{equation}
The linear force--flux relation thus arises from the existence of a 
quadratic dissipation potential.

This potential structure offers a unified interpretation of the 
thermodynamic admissibility conditions established 
in Section~\ref{sec:entropyprod}. Because $\Phi$ is a smooth scalar 
function, the equality of mixed partial derivatives,
\begin{equation}
\frac{\partial^2 \Phi}{\partial X_a \partial X_b}
=
\frac{\partial^2 \Phi}{\partial X_b \partial X_a},
\end{equation}
directly yields the Onsager reciprocal relations $\mathbf{L}_{ab} = 
\mathbf{L}_{ba}^{\mathrm{T}}$ without invoking microscopic 
time-reversibility as a separate postulate. The entropy production 
rate is then
\begin{equation}
\dot{s}_{\mathrm{prod}}
=
\mathbf{J}^{\mathrm{T}}\mathbf{X}
=
\mathbf{X}^{\mathrm{T}}\mathbb{L}\,\mathbf{X}
=
2\,\Phi(\mathbf{X}) \geq 0,
\end{equation}
so that non-negativity of $\Phi$ is equivalent to the second law and 
requires $\mathbb{L}$ to be positive semi-definite. The reciprocity 
and positive semi-definiteness conditions, previously derived as 
separate constraints, both emerge from the single assumption that a 
quadratic dissipation potential exists.

Material symmetry further constrains the admissible structure of
$\mathbb{L}$. For an isotropic medium, each tensor block reduces to a
scalar multiple of the identity tensor,
\begin{equation}
\mathbf{L}_{ab}=L_{ab}\mathbf{I},
\end{equation}
where $\mathbf{I}$ denotes the identity operator of appropriate rank.
Consequently, no preferred direction exists in the constitutive
response, and all directional dependence is carried by the gradient
operators acting on the driving forces, while the magnitude of each
coupling is encoded in the scalar coefficient $L_{ab}$.

The quadratic dissipation potential thus provides a compact and 
systematic generator for thermodynamically admissible multiphysical 
coupling within the linear response regime.

\section{Specialization to Electro--Chemo--Mechanics of Porous Electrodes}
\label{sec: Porous Eletrodes Model}

\subsection{Physical Assumptions}
\label{sec: physical assumption}

We now specialize the general framework developed in 
Section~\ref{sec: General Framework} to porous electrodes operating 
through ion intercalation. Intercalation refers to the reversible 
insertion of guest ions or molecules into a host material with a 
layered or open-framework crystal structure, without disrupting 
the overall structural 
integrity of the host~\citep{whittingham_intercalation_1982}. It is the 
dominant charge-storage mechanism in modern lithium-ion batteries, 
with representative host materials including graphite, lithium iron 
phosphate (LFP), lithium nickel manganese cobalt oxide (NMC), and 
silicon-based alloys. The following physical assumptions are adopted 
to isolate the dominant couplings relevant to battery operation while 
preserving the essential multiphysical structure:

\begin{itemize}
    \item \textbf{Isothermal conditions.} The system is assumed to 
    operate at a uniform reference temperature, so that thermal 
    gradients and temperature variations are neglected during the 
    perturbation. This assumption is consistent with typical 
    laboratory measurements performed in temperature-controlled 
    environments and removes the thermal field from the active set 
    of coupled processes.
    
    \item \textbf{Non-piezoelectric solid matrix.} The active 
    materials are assumed to exhibit no direct piezoelectric coupling 
    between electric field and mechanical stress. This is appropriate 
    for the centrosymmetric crystal structures of most common 
    intercalation electrodes, including layered oxides, spinels, and 
    olivines~\citep{safari_piezoelectricity_2008,park_induced_2022}.
    
    \item \textbf{Migration-diffusion ionic transport.} Ionic transport 
    in the electrolyte is governed by the combined action of 
    electromigration and concentration-driven diffusion. Electrokinetic 
    couplings between electric fields and pore-fluid flow (electroosmosis, 
    streaming potential) are neglected at the homogenized porous-electrode 
    scale, which is justified for the moderate ionic strengths and 
    macroscopic pore sizes characteristic of battery 
    electrolytes~\citep{qu_charging_2018,zhao_electrolytes_nodate}.
    
    \item \textbf{Linearized poro-mechanical response.} The porous 
    electrode is described by a linearized poro-mechanical 
    constitutive law in which pore pressure contributes additively to 
    the macroscopic stress through the Biot effective stress relation, 
    and pore pressure evolves through fluid storage and Darcy-type 
    hydraulic relaxation~\citep{wang_theory_2017}. Nonlinear effects such as finite-deformation 
    poroelasticity, pore collapse, and capillary phenomena are not 
    considered.
\end{itemize}

Under these assumptions, all direct electric--mechanical 
($\mathbf{L}_{13}$, $\mathbf{L}_{31}$) and electric--hydraulic 
($\mathbf{L}_{14}$, $\mathbf{L}_{41}$) couplings vanish, and the 
thermal channel ($\mathbf{L}_{i5}$, $\mathbf{L}_{5j}$) is removed 
from the active matrix. The remaining multiphysical interactions are 
organized along three pathways: electrochemical 
($\mathbf{L}_{12} = \mathbf{L}_{21}^{\mathrm{T}}$), chemo-mechanical 
($\mathbf{L}_{23} = \mathbf{L}_{32}^{\mathrm{T}}$), and poro-mechanical 
($\mathbf{L}_{34} = \mathbf{L}_{43}^{\mathrm{T}}$). These three 
surviving couplings define the irreducible multiphysical structure of 
the porous electrode within the present framework.

\subsection{Reduced Constitutive Matrix}
\label{sec: Reduced Onsager Structure}

For the one-dimensional through-thickness analysis adopted in 
subsequent sections, the tensorial flux--force relations of 
Section~\ref{sec: General Framework} reduce to scalar relations along 
the stacking direction. The generalized flux and driving-force vectors 
become
\begin{equation}
\mathbf{J} = 
\begin{pmatrix}
J_e \\
F_i \\
\dot{\varepsilon}^p \\
Q
\end{pmatrix},
\qquad
\mathbf{X} = 
\begin{pmatrix}
-\nabla \phi \\
-\nabla \mu \\
-\sigma^p \\
-\nabla p
\end{pmatrix},
\end{equation}
where $J_e$, $F_i$, $\dot{\varepsilon}^p$, and $Q$ denote, respectively, 
the electric current density, ionic flux, volumetric strain rate of the 
porous medium, and pore-fluid flux along the through-thickness direction. 
The thermal channel has been removed under the isothermal assumption 
of Section~\ref{sec: physical assumption}.

Invoking the simplifications introduced in 
Section~\ref{sec: physical assumption}---the absence of piezoelectric 
coupling ($L_{13} = L_{31} = 0$), the absence of electrokinetic 
coupling ($L_{14} = L_{41} = 0$), and the absence of direct 
ion--fluid advective coupling ($L_{24} = L_{42} = 0$)---the 
constitutive matrix takes the reduced block-tridiagonal form
\begin{equation}
\begin{pmatrix}
J_e \\
F_i \\
\dot{\varepsilon}^p \\
Q
\end{pmatrix}
=
\begin{pmatrix}
L_{11} & L_{12} & 0        & 0       \\
L_{12} & L_{22} & L_{23} & 0       \\
0         & L_{23} & L_{33} & L_{34} \\
0         & 0         & L_{34} & L_{44}
\end{pmatrix}
\begin{pmatrix}
-\nabla \phi \\
-\nabla \mu \\
-\sigma^p \\
-\nabla p
\end{pmatrix},
\label{eq:reduced_onsager_matrix}
\end{equation}
where Onsager reciprocity has been imposed to set 
$L_{ij} = L_{ji}$. The reduced matrix preserves only three 
multiphysical couplings: the electrochemical coupling $L_{12}$ 
between ionic transport and electric current, the chemo-mechanical 
coupling $L_{23}$ between intercalation and deformation, and the 
poro-mechanical coupling $L_{34}$ between skeletal deformation and 
pore-fluid flow.

The block-tridiagonal structure of 
Eq.~\ref{eq:reduced_onsager_matrix} reveals an important physical 
feature: the four transport processes form a sequential coupling chain. 
Electric fields drive ionic transport ($L_{12}$), which in turn 
generates intercalation-induced deformation ($L_{23}$), which 
finally drives pore-fluid redistribution through poroelastic coupling 
($L_{34}$). There is no direct pathway from electric current to 
mechanical or hydraulic response; the mechanical signal probed by MEIS 
arises only through the cascade of these three reciprocal couplings. 
This sequential structure underlies the multiplicative form of the 
MEIS transfer function derived in subsequent sections.

\paragraph{Thermodynamic admissibility}

The reduced Onsager matrix must satisfy the second law of thermodynamics.
The local rate of entropy production is

\begin{equation}
\mathcal{D}
=
J_e (-\nabla \phi)
+
F_i (-\nabla \mu)
+
\dot{\varepsilon}\,\sigma
+
Q (-\nabla p),
\end{equation}

which can be written compactly as
\begin{equation}
\mathcal{D}
=
\mathbf{X}^{\mathrm{T}}
\mathbf{L}
\mathbf{X}.
\end{equation}

Thermodynamic admissibility requires
\[
\mathcal{D} \ge 0
\quad \text{for all admissible } \mathbf{X},
\]
which implies that the constitutive matrix $\mathbf{L}$ must be symmetric and
positive semi-definite. In particular, the diagonal coefficients must
satisfy
\[
L_{11}>0, \qquad
L_{22}>0, \qquad
L_{33}>0, \qquad
L_{44}>0,
\]
while the principal minors impose bounds on the strength of the coupling
coefficients.

\subsection{Governing Field Equations}

The reduced constitutive relations of 
Section~\ref{sec: Reduced Onsager Structure} must be combined with 
the conservation laws governing each field to obtain a closed system 
of governing equations. The unknown fields are the electric potential 
$\phi$, the ionic concentration $c$, the mechanical strain $\varepsilon$, 
and the pore pressure $p$, all taken as functions of the through-thickness 
coordinate $x$ and time $t$. Under the one-dimensional approximation 
adopted throughout this section, the governing equations take the 
following form.

\paragraph{Charge conservation}
Electric charge conservation in the absence of bulk charge accumulation 
requires
\begin{equation}
\partial_x J_e = 0,
\label{eq:charge_conservation}
\end{equation}
so that the electric current density is spatially uniform across the 
electrode thickness.

\paragraph{Species conservation}
Conservation of the inserted ionic species in the active material is 
expressed as
\begin{equation}
\partial_t c + \partial_x F_i = 0,
\label{eq:species_conservation}
\end{equation}
where $c$ is the volumetric concentration of intercalated species and 
$F_i$ is the ionic flux.

\paragraph{Mechanical equilibrium}
Under quasi-static loading at the time scales of interest, the 
porous-medium stress satisfies the equilibrium equation
\begin{equation}
\partial_x \sigma^p = 0,
\label{eq:mechanical_equilibrium}
\end{equation}
so that $\sigma^p$ is spatially uniform across the electrode thickness. 
At the homogenized porous-medium scale, $\sigma^p$ is the total stress 
that, in the presence of a pore fluid, decomposes through the Biot 
effective stress relation~\citep{biot_general_1941} into a skeleton 
contribution and a pore-pressure contribution,
\begin{equation}
\sigma^p = \sigma^s - \alpha\,p,
\label{eq:effective_stress}
\end{equation}
where $\sigma^s$ is the skeleton stress and $\alpha$ is the Biot 
coefficient.

\paragraph{Fluid mass conservation}
Conservation of pore fluid takes the form
\begin{equation}
\partial_t \zeta + \partial_x Q = 0,
\label{eq:fluid_conservation}
\end{equation}
where $\zeta$ is the increment of fluid content (fluid volume gained 
per unit reference volume) and $Q$ is the pore-fluid flux. The evolution 
of $\zeta$ couples to the deformation of the solid skeleton and to the 
pore pressure through the constitutive structure, which will be specified 
together with the poro-mechanical closure in the following sections.

\bigskip

Together with the reduced constitutive relations 
[Eq.~\ref{eq:reduced_onsager_matrix}], 
Eqs.~\ref{eq:charge_conservation}--\ref{eq:fluid_conservation} form 
a closed system describing the coupled electrochemical, mechanical, 
and hydraulic response of the porous intercalation electrode.

\section{Current-Controlled Mechano-Electrochemical Impedance Spectroscopy}

\subsection{Concept and Motivation}

Electrochemical systems based on porous electrodes inherently couple 
charge transport, species redistribution, solid deformation, and fluid 
flow. Conventional electrochemical impedance spectroscopy 
(EIS)~\citep{wang_electrochemical_2021} probes the electrical response 
to a small harmonic current perturbation, thereby characterizing 
interfacial kinetics and transport in the electrochemical domain. 
However, electrochemical processes also generate chemical strain and 
internal stress, which encode multiphysical information that is 
invisible in purely electrical measurements. The mechanical counterpart 
to EIS is dynamic mechanical analysis 
(DMA)~\citep{menard2020dynamic,ferry1961viscoelastic}, in which a 
small harmonic strain (or displacement) is imposed and the resulting 
stress (or force) is recorded. DMA characterizes the viscoelastic 
response of the material but provides no information about the 
underlying electrochemical processes. To bridge these two complementary 
techniques, our group recently developed mechano-electrochemical 
impedance spectroscopy (MEIS)~\citep{fang_mechano-electrochemical_2025}, 
which extends the impedance concept to the coupled 
electro-chemo-mechanical regime. In MEIS, a small sinusoidal electric 
current is applied to the electrode and the resulting linear mechanical 
response---stress or pressure---is measured. The frequency-domain 
transfer function between current and mechanical response captures 
the dynamics of electrochemical, chemical, and mechanical processes 
on an equal footing.

The reduced constitutive matrix derived in 
Section~\ref{sec: Reduced Onsager Structure} provides the local 
relations between fluxes and driving forces underlying this coupled 
response. In the following subsections, we combine these constitutive 
relations with the corresponding balance laws in the frequency domain 
to derive the analytical MEIS transfer function. The resulting 
expression decomposes naturally into physically interpretable 
contributions associated with electrochemical accumulation, 
viscoelastic relaxation, hydraulic relaxation, and porosity 
accommodation, each carrying a distinct characteristic time scale. 
Because the perturbation amplitude is small, the response remains 
within the linear regime: superposition applies, and the distinct 
multiphysical relaxation processes separate cleanly in frequency 
space. MEIS thus provides a frequency-resolved probe of 
electro-chemo-mechanical coupling in porous electrodes.

\subsection{Problem Definition}

We consider a one-dimensional porous electrode of characteristic 
thickness $\ell$, operated near a local equilibrium state. The 
governing equations introduced in 
Section~\ref{sec: Porous Eletrodes Model} describe the coupled evolution 
of electric potential $\phi$, chemical potential $\mu$, concentration 
$c$ of the inserted species, mechanical deformation of the porous 
skeleton, and pore pressure $p$.

In a current-controlled MEIS experiment, a small harmonic electronic 
current density is imposed at the electrode boundary:
\begin{equation}
J_e(t) = \hat{J}_e\, e^{i\omega t},
\label{eq: MEIS current}
\end{equation}
where $\omega$ is the angular frequency, $i$ is the imaginary unit, and 
$\hat{J}_e$ is the complex amplitude of the applied current. Because 
the perturbation amplitude is small, the system response remains within 
the linear regime, and all state variables oscillate harmonically at 
the same frequency $\omega$. The electrode stack is mechanically 
constrained under global displacement control, so that the macroscopic 
strain vanishes,
\begin{equation}
\varepsilon = 0.
\label{eq: zero strain}
\end{equation}
Local stresses, pore pressure, and porosity evolution within the 
representative porous microstructure are unconstrained and develop in 
response to the applied current.

Since the system is linear and the excitation is purely harmonic, we 
formulate the problem in the frequency domain via the temporal Fourier 
transform. For any time-dependent field $f(x, t)$, we define
\begin{equation}
\hat{f}(x, \omega) = \int_{-\infty}^{\infty} f(x, t)\, e^{-i\omega t}\, \mathrm{d}t,
\label{eq: fourier transform}
\end{equation}
so that time derivatives transform according to 
$\partial_t \;\rightarrow\; i\omega$. In the linear regime, each field 
variable is fully characterized by its complex amplitude $\hat{f}(x, 
\omega)$ at the excitation frequency, and all governing equations 
become algebraic relations between these amplitudes.

The mechano-electrochemical impedance is then defined as the 
frequency-domain transfer function between the applied current and 
the resulting macroscopic stress response,
\begin{equation}
Z_{\mathrm{MEIS}}(\omega)
= \frac{\hat{\sigma}_{\mathrm{tot}}(\omega)}{\hat{J}_e(\omega)},
\label{eq: MEIS definition}
\end{equation}
where $\hat{\sigma}_{\mathrm{tot}}$ is the complex amplitude of the 
total stress measured by the load cell at the electrode boundary. The 
following subsections combine the reduced constitutive 
relation matrix with the corresponding balance laws to evaluate this 
transfer function in closed form.

\subsection{Derivation of the General MEIS Expression}
\label{sec: MEIS derivation}

We now combine the reduced constitutive relations of
Section~\ref{sec: Reduced Onsager Structure} with the balance laws in
the frequency domain to obtain the MEIS transfer function. The
derivation separates naturally into three stages that follow the
sequential coupling chain of the reduced matrix: the applied current
drives ionic accumulation (electrochemical), the accumulated species
generate skeleton stress and pore pressure (chemo- and poro-mechanical),
and these combine into the measured impedance. The complete
step-by-step algebra, retaining all couplings and relaxation time
scales, is given in Appendix~\ref{app: full derivation}; here we
establish the structure and the resulting general expression. Throughout,
the perturbation is harmonic, $\hat f(t)=\hat f\,e^{i\omega t}$, so
$\partial_t\to i\omega$, and the electrode is treated in the lumped
(thickness-averaged) limit.

\paragraph{Stage 1: electrochemical accumulation}
Charge conservation renders the electronic current uniform, and
eliminating the electric-potential gradient from the first two rows of
the reduced matrix expresses the ionic flux in terms of the applied
current through the cation transference number
\begin{equation}
t_+ \equiv \mathrm{F}_{\!c}\,\frac{L_{12}}{L_{11}},
\label{eq:transference_definition}
\end{equation}
the fraction of the ionic current carried by the inserted cation, where $\mathrm{F}_{\!c}$ is the Faraday constant (96 485.33 C/mol). The
thickness-integrated species balance---accumulation equal to the
Faradaic flux entering at the separator---then gives the concentration
response
\begin{equation}
\hat c(\omega) = \mathcal{G}_e\,\mathcal{G}_c(\omega)\,\hat J_e,
\qquad
\mathcal{G}_e = \frac{t_+}{\mathrm{F}_{\!c}},
\qquad
\mathcal{G}_c(\omega) = \frac{1}{\ell\,i\omega},
\label{eq:concentration_factored}
\end{equation}
where $\mathcal{G}_e$ is the electrical-to-ionic conversion factor and
$\mathcal{G}_c$ the Faradaic accumulation factor. The $1/(i\omega)$ in
$\mathcal{G}_c$ is the integrator that converts current into stored
concentration, producing the capacitor-like signature familiar from EIS.
Diffusion and chemo-mechanical feedback generalize $\mathcal{G}_c$ to a
full relaxation denominator $\Omega^{*}(\omega)$
[Appendix~\ref{app: full derivation}, Eq.~\ref{eq:app_Omega_star}];
these are negligible over the frequency range of interest and are
omitted in the present expression.

\paragraph{Stage 2: mechanical response}
The accumulated species produce a chemical eigenstrain that, under the
global constraint $\varepsilon=0$ imposed by the MEIS fixture, generates
an internal stress through two distinct pathways. Through the
chemo-mechanical coupling, the eigenstrain loads the viscoelastic solid
skeleton; through the poro-mechanical coupling, the accompanying
pore-volume change pressurizes the pore fluid. Both responses are linear
in $\hat c$, so the total stress measured at the boundary may be written
\begin{equation}
\hat\sigma_{\mathrm{tot}}(\omega)
= \bigl[\mathcal{H}_{cm}(\omega) + \mathcal{H}_{pm}(\omega)\bigr]\,\hat c,
\label{eq:Hcm_Hpm_definitions}
\end{equation}
which defines the \emph{chemo-mechanical kernel} $\mathcal{H}_{cm}$
(concentration to skeleton stress) and the \emph{poro-mechanical kernel}
$\mathcal{H}_{pm}$ (concentration to the pore-fluid contribution to
stress). The two kernels enter additively because the Biot
effective-stress relation superposes the skeleton and pore-fluid
stresses linearly. Their explicit forms---carrying the viscoelastic,
porosity-accommodation, and hydraulic relaxation time scales---are
derived in Section~\ref{sec: reduced model}.

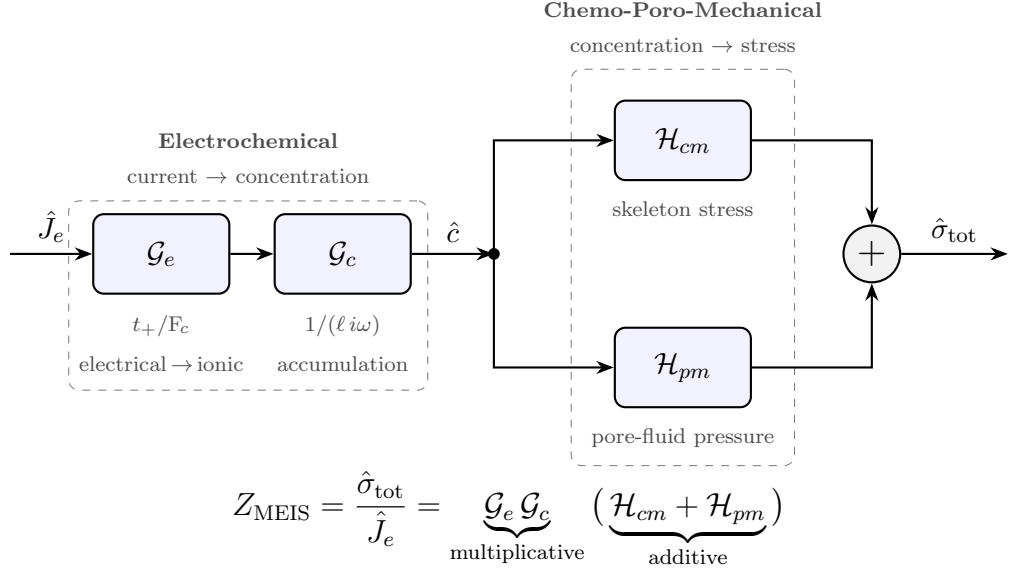
\begin{figure}[t]
\centering
\begin{tikzpicture}[>=Stealth, font=\small,
  block/.style={draw,rounded corners,minimum width=1.8cm,minimum height=1.05cm,
                align=center,thick,fill=blue!5},
  sum/.style={draw,circle,thick,minimum size=0.75cm,inner sep=0pt,fill=gray!10},
  desc/.style={font=\scriptsize,align=center,text=black!70},
  grp/.style={draw,dashed,rounded corners,gray,inner sep=0.40cm}]

\coordinate (Jin) at (-2.0,0);
\node[block] (Ge) at (0,0)    {$\mathcal{G}_e$};
\node[block] (Gc) at (2.4,0)  {$\mathcal{G}_c$};
\coordinate (cn) at (4.4,0);
\node[block] (Hcm) at (6.9,1.5)  {$\mathcal{H}_{cm}$};
\node[block] (Hpm) at (6.9,-1.5) {$\mathcal{H}_{pm}$};
\node[sum]   (sum) at (9.4,0) {\large$+$};
\coordinate (out) at (11.2,0);

\node[desc,below=0.12cm of Ge]  (dGe)  {$t_+/\mathrm{F}_{\!c}$\\[1pt]electrical$\,\to\,$ionic};
\node[desc,below=0.12cm of Gc]  (dGc)  {$1/(\ell\,i\omega)$\\[1pt]accumulation};
\node[desc,below=0.12cm of Hcm] (dHcm) {skeleton stress};
\node[desc,below=0.12cm of Hpm] (dHpm) {pore-fluid pressure};

\begin{scope}[on background layer]
  \node[
  draw,dashed,rounded corners,gray,
  minimum width=4.75cm,
  minimum height=2.5cm
] (gE) at (1.15,-0.55) {};
  
  \coordinate (gHbottom) at ($(dHpm.south)+(0,0.3cm)$);
  \node[grp,fit=(Hcm)(Hpm)(dHcm)(gHbottom)] (gH) {};
\end{scope}
\node[desc,above=0.08cm of gE] {\textbf{Electrochemical}\\[-1pt]current $\to$ concentration};
\node[desc,above=0.08cm of gH] {\textbf{Chemo-Poro-Mechanical}\\[-1pt]concentration $\to$ stress};

\draw[->,thick] (Jin) -- node[above]{$\hat J_e$} (Ge);
\draw[->,thick] (Ge) -- (Gc);
\draw[->,thick] (Gc) -- node[above]{$\hat c$} (cn);
\fill (cn) circle (2.2pt);
\draw[->,thick] (cn) |- (Hcm);
\draw[->,thick] (cn) |- (Hpm);
\draw[->,thick] (Hcm) -| (sum);
\draw[->,thick] (Hpm) -| (sum);
\draw[->,thick] (sum) -- node[above]{$\hat\sigma_{\mathrm{tot}}$} (out);

\node[font=\normalsize] at (4.6,-3.5)
  {$\displaystyle Z_{\mathrm{MEIS}}=\frac{\hat\sigma_{\mathrm{tot}}}{\hat J_e}
     =\underbrace{\mathcal{G}_e\,\mathcal{G}_c}_{\text{multiplicative}}
      \big(\underbrace{\mathcal{H}_{cm}+\mathcal{H}_{pm}}_{\text{additive}}\big)$};
\end{tikzpicture}
\caption{Block-diagram structure of the MEIS transfer function. The
applied current $\hat J_e$ is converted to an intercalated-species
concentration $\hat c$ by the two electrochemical factors---the
electrical-to-ionic conversion $\mathcal{G}_e=t_+/\mathrm{F}_{\!c}$ and
the Faradaic accumulation $\mathcal{G}_c=1/(\ell\,i\omega)$---which act in
series (multiplicatively). The concentration then drives two parallel
mechanical pathways: the chemo-mechanical kernel $\mathcal{H}_{cm}$
(skeleton stress) and the poro-mechanical kernel $\mathcal{H}_{pm}$
(pore-fluid pressurization), which add through the Biot effective-stress
relation to give the total stress $\hat\sigma_{\mathrm{tot}}$. The
resulting impedance is the multiplicative-plus-additive form
$Z_{\mathrm{MEIS}}=\mathcal{G}_e\mathcal{G}_c(\mathcal{H}_{cm}+\mathcal{H}_{pm})$.}
\label{fig:transfer_structure}
\end{figure}

\paragraph{Stage 3: the transfer function}
Combining Eqs.~\ref{eq:concentration_factored} and
\ref{eq:Hcm_Hpm_definitions}, the mechano-electrochemical impedance is
\begin{equation}
\boxed{\;
Z_{\mathrm{MEIS}}(\omega)
= \frac{\hat\sigma_{\mathrm{tot}}(\omega)}{\hat J_e(\omega)}
= \mathcal{G}_e\,\mathcal{G}_c(\omega)
  \bigl[\mathcal{H}_{cm}(\omega) + \mathcal{H}_{pm}(\omega)\bigr].
\;}
\label{eq:MEIS_general_structure}
\end{equation}
As illustrated in Figure~\ref{fig:transfer_structure},
the result is a product of two electrochemical factors
($\mathcal{G}_e\mathcal{G}_c$, converting current to concentration)
multiplying a sum of two mechanical kernels
($\mathcal{H}_{cm}+\mathcal{H}_{pm}$, converting concentration to stress).
The multiplicative form reflects the sequential coupling chain---current
drives ion flux, which drives concentration accumulation, which in turn
drives both skeleton stress and pore-fluid pressurization---while the
additivity of the mechanical kernels reflects the Biot superposition of
skeleton and fluid contributions. This multiplicative-plus-additive
structure is the essential feature distinguishing MEIS from EIS, in
which all contributions to the impedance enter additively.

\paragraph{A remark on sign}
The continuum stress $\sigma_{\mathrm{tot}}$ is taken
positive in tension (Section~\ref{sec: General Framework}), so that the
constrained chemical expansion places the electrode in compression
($\hat\sigma_{\mathrm{tot}}<0$ for $\hat J_e>0$); with the kernel forms
of Section~\ref{sec: reduced model} this gives a negative real prefactor
for $Z_{\mathrm{MEIS}}$. A load cell instead reports the compressive
stack stress $-\sigma_{\mathrm{tot}}$, and the impedance referred to that
measured quantity carries the opposite, positive sign
[Appendix~\ref{app: full derivation}, Eq.~\ref{eq:app_MEIS_meas}]. The
two differ only by this global sign; the relative magnitudes and phase
relationships among the contributions are unaffected.

\subsection{Constitutive Closure: Chemo-Mechanical and Poro-Mechanical Kernels}
\label{sec: reduced model}

To complete the MEIS expression we specify the two kernels
$\mathcal{H}_{cm}$ and $\mathcal{H}_{pm}$ introduced in
Eq.~\ref{eq:Hcm_Hpm_definitions}. Both arise from the mechanical and
hydraulic constitutive response of the porous electrode, which must be
specified beyond the linear constitutive relations to capture the
rheology and pore-scale physics of real materials. We present the
closure here; the complete step-by-step assembly of the transfer
function is given in Appendix~\ref{app: full derivation}.

\paragraph{Strain decomposition under the global constraint}
The electrode-level strain is the sum of an elastic skeleton strain
$\varepsilon_e$, the chemical (intercalation) eigenstrain $\beta c$ with
$\beta$ the chemical expansion coefficient, and the relative pore
contraction $\xi$ that accommodates part of the particle swelling
through rearrangement of the porous microstructure:
\begin{equation}
\varepsilon = \varepsilon_e + \beta c - \xi.
\label{eq:strain_decomposition}
\end{equation}
Under the global constraint $\varepsilon = 0$ imposed by the MEIS
fixture, the elastic strain is
\begin{equation}
\hat\varepsilon_e = -\beta\hat c + \hat\xi,
\label{eq:elastic_strain_constraint}
\end{equation}
so the chemical eigenstrain is partly relieved by the porosity change
$\hat\xi$ rather than transmitted entirely to the skeleton. 
This strain decomposition is inspired by our recent experimental observation~\citep{fang_mechano-electrochemical_2025}.

\paragraph{Skeleton viscoelasticity}
Real porous electrodes exhibit both an instantaneous elastic and a
time-dependent viscous response. We represent the skeleton by a standard
linear solid (SLS)---a spring $E_\infty$ in parallel with a Maxwell
element (spring $E_1$ in series with a dashpot)---giving the complex
modulus
\begin{equation}
\hat\sigma^s = E^*(\omega)\,\hat\varepsilon_e,
\qquad
E^*(\omega) = \frac{E_\infty + i\omega\tau_m E_0}{1 + i\omega\tau_m},
\label{eq:SLS_modulus}
\end{equation}
where $E_0 = E_\infty + E_1$ is the instantaneous (unrelaxed) modulus,
$E_\infty$ the relaxed modulus, and $\tau_m$ the viscoelastic relaxation
time.

\begin{figure}[t]
\centering
\begin{tikzpicture}[>=Stealth, font=\small,
  box/.style={draw,rounded corners,thick,fill=blue!5},
  desc/.style={font=\scriptsize,align=center,text=black!70},
  lab/.style={font=\scriptsize}]

\node[box,minimum width=2.9cm,minimum height=1.6cm] (A) at (2.5,0) {};
\node[box,minimum width=3.8cm,minimum height=1.6cm] (B) at (8.0,0) {};

\draw[->,thick] (-0.9,0) -- node[above]{$\beta\hat c$} (A.west);
\draw[->,thick] (A.east) -- node[above,lab]{$\hat\varepsilon_e=-\beta\mathcal{B}\hat c$} (B.west);
\draw[->,thick] (B.east) -- node[above]{$\hat\sigma^s$} (11.7,0);

\draw[thick] (1.35,0)--(1.55,0);   \draw[thick] (3.45,0)--(3.65,0);
\draw[thick] (1.55,-0.32)--(1.55,0.32);  \draw[thick] (3.45,-0.32)--(3.45,0.32);
\hspring{1.55}{3.45}{0.32}
\hdashpot{1.55}{3.45}{-0.32}
\node[lab] at (2.5,0.60) {$k$};
\node[lab] at (2.5,-0.66) {$\eta_\xi$};

\draw[thick] (6.40,0)--(6.65,0);   \draw[thick] (9.35,0)--(9.60,0);
\draw[thick] (6.65,-0.34)--(6.65,0.34);  \draw[thick] (9.35,-0.34)--(9.35,0.34);
\hspring{6.65}{9.35}{0.34}
\hspring{6.65}{7.95}{-0.34}
\hdashpot{7.95}{9.35}{-0.34}
\node[lab] at (8.0,0.62) {$E_\infty$};
\node[lab] at (7.25,-0.66) {$E_1$};
\node[lab] at (8.70,-0.66) {$\eta_m$};

\node[desc,above=0.12cm of A] {\textbf{Porosity accommodation}};
\node[desc,above=0.12cm of B] {\textbf{Skeleton viscoelasticity} (SLS)};
\node[desc,below=0.12cm of A] {bridge $\mathcal{B}(\omega)$\\[1pt]$\tau_\xi=\eta_\xi/(E^*{+}k)$};
\node[desc,below=0.12cm of B] {modulus $E^*(\omega)$\\[1pt]$\tau_m=\eta_m/E_1$};
\node[desc] at (-0.9,-0.5) {chemical\\eigenstrain};
\node[desc] at (11.0,-0.5) {skeleton\\stress};

\node[font=\normalsize] at (5.2,-2.55)
  {$\displaystyle \mathcal{H}_{cm}(\omega)=-\beta\,
     \underbrace{E^*(\omega)}_{\text{viscoelastic}}\,
     \underbrace{\mathcal{B}(\omega)}_{\text{accommodation}}$};
\end{tikzpicture}
\caption{Rheological structure of the chemo-mechanical kernel
$\mathcal{H}_{cm}$. The chemical eigenstrain $\beta\hat c$ generated by
intercalation is processed by two relaxation mechanisms in series. The
\emph{porosity accommodation} stage (a Kelvin--Voigt element with
microstructural stiffness $k$ and viscosity $\eta_\xi$) lets the
pore network rearrange and absorb a fraction $\xi_0$ of the swelling,
reducing the strain transmitted to the skeleton to
$\hat\varepsilon_e=-\beta\mathcal{B}(\omega)\hat c$ with relaxation time
$\tau_\xi$. The \emph{skeleton viscoelasticity} stage---a standard
linear solid (a spring $E_\infty$ in parallel with a Maxwell arm
$E_1$--$\eta_m$)---then converts this elastic strain into skeleton stress
through the complex modulus $E^*(\omega)$ with relaxation time $\tau_m$.
At low frequency the pores rearrange ($\mathcal{B}\to1-\xi_0$) and the
skeleton relaxes ($E^*\to E_\infty$); at high frequency the porosity is
frozen ($\mathcal{B}\to1$) and the skeleton is stiff ($E^*\to E_0$). The
two combine as $\mathcal{H}_{cm}=-\beta\,E^*(\omega)\,\mathcal{B}(\omega)$.}
\label{fig:chemo_mechanical}
\end{figure}
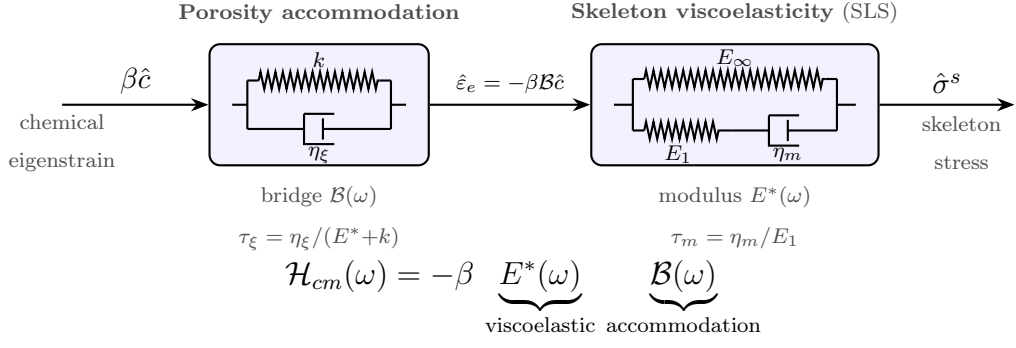

\paragraph{Porosity evolution from a free-energy functional}
To place the porosity change $\xi$ on a thermodynamic footing---rather
than introduce it as a phenomenological fit---we posit a Helmholtz
free-energy density
\begin{equation}
\psi = \frac12\,E^*\varepsilon_e^{\,2} + \frac12\,k\,\xi^{2},
\label{eq:free_energy_porosity}
\end{equation}
in which $k$ is the microstructural stiffness resisting
consolidation of the granular packing and binder network. The
thermodynamic force conjugate to $\xi$ is
$f_\xi = -\partial\psi/\partial\xi|_{\varepsilon,c}$, which with
Eq.~\ref{eq:elastic_strain_constraint} evaluates to
$f_\xi = E^*\beta c - (E^*+k)\xi$. A linear kinetic law
$\dot\xi = f_\xi/\eta_\xi$, with $\eta_\xi$ the microstructural viscosity
governing dissipation during particle rearrangement and binder creep,
yields the first-order relaxation
\begin{equation}
\dot\xi + \frac{\xi}{\tau_\xi} = \frac{E^*\beta}{\eta_\xi}\,c,
\qquad
\tau_\xi = \frac{\eta_\xi}{E^*+k},
\label{eq:porosity_evolution}
\end{equation}
whose frequency-domain solution is
\begin{equation}
\hat\xi = \frac{\xi_0\,\beta}{1+i\omega\tau_\xi}\,\hat c,
\qquad
\xi_0 = \frac{E^*}{E^*+k},
\label{eq:porosity_response}
\end{equation}
with $\xi_0$ the equilibrium porosity-accommodation ratio, defined at the moment when the spectroscopy test is carried out. Substituting
into Eq.~\ref{eq:elastic_strain_constraint} gives
$\hat\varepsilon_e = -\beta\,\mathcal{B}(\omega)\,\hat c$, where
\begin{equation}
\mathcal{B}(\omega) = 1 - \frac{\xi_0}{1+i\omega\tau_\xi}
= \frac{(1-\xi_0) + i\omega\tau_\xi}{1+i\omega\tau_\xi}
\label{eq:bridge_function}
\end{equation}
is the bridge function mapping particle-scale chemical strain to
electrode-scale elastic strain. 
It interpolates between 1)
$\mathcal{B}\to 1-\xi_0 = k/(E^*+k)$ at low frequency---porosity
rearranges and absorbs part of the expansion---and 2) $\mathcal{B}\to1$ at
high frequency, where the porosity is frozen and the full chemical
eigenstrain is transmitted to elastic strain.
Equation~\ref{eq:bridge_function} thus supplies the thermodynamic origin,
in terms of the microstructural parameters $k$ and $\eta_\xi$, of
the bridge function introduced phenomenologically by
\citet{fang_mechano-electrochemical_2025}.

\paragraph{Chemo-mechanical kernel}
Combining Eqs.~\ref{eq:SLS_modulus} and \ref{eq:bridge_function}, the
skeleton stress is
$\hat\sigma^s = -\beta\,E^*(\omega)\,\mathcal{B}(\omega)\,\hat c$, so the
chemo-mechanical kernel is
\begin{equation}
\mathcal{H}_{cm}(\omega) = -\beta\,E^*(\omega)\,\mathcal{B}(\omega).
\label{eq:Hcm_specialized}
\end{equation}

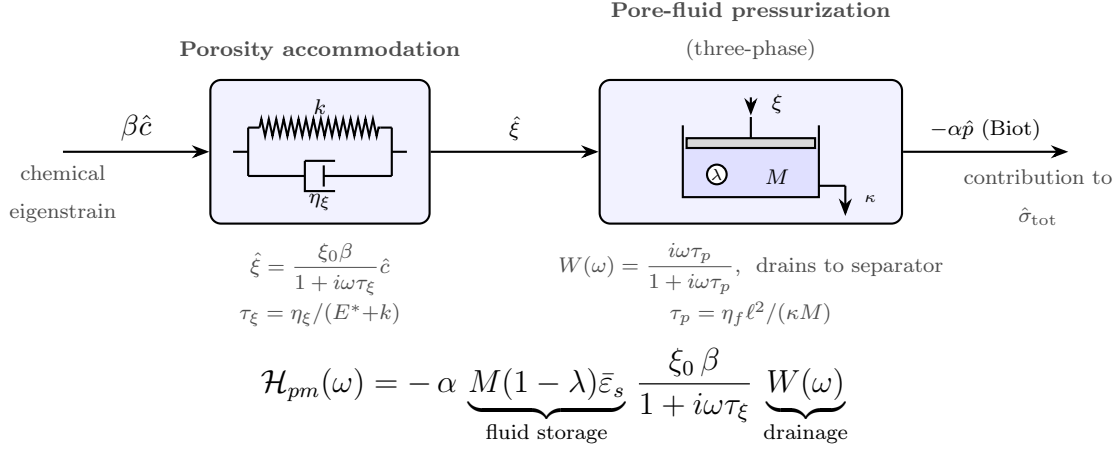
\begin{figure}[t]
\centering
\begin{tikzpicture}[>=Stealth, font=\small,
  box/.style={draw,rounded corners,thick,fill=blue!5},
  desc/.style={font=\scriptsize,align=center,text=black!70},
  lab/.style={font=\scriptsize}]

\node[box,minimum width=2.9cm,minimum height=1.9cm] (A) at (2.5,0) {};
\node[box,minimum width=4.0cm,minimum height=1.9cm] (B) at (8.2,0) {};

\draw[->,thick] (-0.9,0) -- node[above]{$\beta\hat c$} (A.west);
\draw[->,thick] (A.east) -- node[above,lab]{$\hat\xi$} (B.west);
\draw[->,thick] (B.east) -- node[above,lab]{$-\alpha\hat p$ (Biot)} (12.4,0);

\draw[thick] (1.35,0)--(1.55,0);   \draw[thick] (3.45,0)--(3.65,0);
\draw[thick] (1.55,-0.32)--(1.55,0.32);  \draw[thick] (3.45,-0.32)--(3.45,0.32);
\hspring{1.55}{3.45}{0.32}
\hdashpot{1.55}{3.45}{-0.32}
\node[lab] at (2.5,0.60) {$k$};
\node[lab] at (2.5,-0.66) {$\eta_\xi$};

\fill[blue!12] (7.3,-0.6) rectangle (9.1,0.05);
\draw[thick] (7.3,0.35)--(7.3,-0.6)--(9.1,-0.6)--(9.1,0.35);
\fill[white] (7.75,-0.30) circle (0.13); \draw[thick] (7.75,-0.30) circle (0.13);
\node[font=\tiny] at (7.75,-0.30) {$\lambda$};
\node[lab] at (8.55,-0.33) {$M$};
\fill[gray!30] (7.35,0.05) rectangle (9.05,0.18);
\draw[thick] (7.35,0.05) rectangle (9.05,0.18);
\draw[thick] (8.2,0.18)--(8.2,0.46);
\draw[->,thick] (8.2,0.72)--(8.2,0.48);
\node[lab] at (8.55,0.64) {$\xi$};
\draw[thick] (9.1,-0.45)--(9.45,-0.45);
\draw[->,thick] (9.45,-0.45)--(9.45,-0.85);
\node[font=\tiny] at (9.78,-0.62) {$\kappa$};

\node[desc,above=0.10cm of A] {\textbf{Porosity accommodation}};
\node[desc,above=0.10cm of B] {\textbf{Pore-fluid pressurization}\\(three-phase)};
\node[desc,below=0.10cm of A] {$\hat\xi=\dfrac{\xi_0\beta}{1+i\omega\tau_\xi}\hat c$\\[2pt]$\tau_\xi=\eta_\xi/(E^*{+}k)$};
\node[desc,below=0.10cm of B] {$W(\omega)=\dfrac{i\omega\tau_p}{1+i\omega\tau_p}$,\ \ drains to separator\\[2pt]$\tau_p=\eta_f\ell^2/(\kappa M)$};
\node[desc] at (-0.9,-0.55) {chemical\\eigenstrain};
\node[desc] at (12.0,-0.6) {contribution to\\$\hat\sigma_{\mathrm{tot}}$};

\node[font=\normalsize] at (5.6,-3.25)
  {$\displaystyle \mathcal{H}_{pm}(\omega)=
     -\,\alpha\,\underbrace{M(1-\lambda)\bar\varepsilon_s}_{\text{fluid storage}}\,
     \frac{\xi_0\,\beta}{1+i\omega\tau_\xi}\,
     \underbrace{W(\omega)}_{\text{drainage}}$};
\end{tikzpicture}
\caption{Structure of the poro-mechanical kernel $\mathcal{H}_{pm}$,
which shares the porosity-accommodation stage of $\mathcal{H}_{cm}$ but
routes the porosity change into the pore fluid rather than the solid
skeleton. The chemical eigenstrain $\beta\hat c$ first produces the pore
contraction $\hat\xi=\xi_0\beta\hat c/(1+i\omega\tau_\xi)$
(relaxation time $\tau_\xi$). In the three-phase pore space, a fraction
$(1-\lambda)$ of this contraction displaces electrolyte while a fraction
$\lambda$ closes gas-filled void without pressurization; the displaced
fluid (storage modulus $M$, solid fraction $\bar\varepsilon_s$) loads the
pore pressure and drains by Darcy flow (permeability $\kappa$) to the
separator, giving the hydraulic relaxation $W(\omega)$ with time
$\tau_p$. Pore pressure builds up under fast compression ($W\to1$,
undrained) and vents at low frequency ($W\to0$, drained). The pressure
enters the total stress through the Biot coefficient $\alpha$, yielding
$\mathcal{H}_{pm}=-\alpha M(1-\lambda)\bar\varepsilon_s\,\xi_0\beta\,
W(\omega)/(1+i\omega\tau_\xi)$.}
\label{fig:poro_mechanical}
\end{figure}

\paragraph{Three-phase poro-mechanical closure}
The pore pressure is driven by displacement of pore fluid as the
porosity rearranges. In a real electrode the pore space holds both
electrolyte and gas-filled void regions, so only part of the porosity
change displaces fluid. We introduce the void accommodation fraction
$\lambda\in[0,1]$, defined such that a fraction $(1-\lambda)$ of the
porosity change displaces fluid while a fraction $\lambda$ closes void
space without generating pressure; the Biot fluid-content increment is
\begin{equation}
\hat\zeta = -(1-\lambda)\,\bar\varepsilon_s\,\hat\xi + \frac{\hat p}{M},
\label{eq:fluid_storage}
\end{equation}
with $\bar\varepsilon_s$ the steady-state solid volume fraction and $M$
the Biot modulus. Combining fluid mass conservation with Darcy flow in
the lumped (thickness-averaged) limit---the drained-separator,
sealed-collector balance derived in Appendix~\ref{app: full derivation}
(Step~4)---gives the first-order pore-pressure relaxation
\begin{equation}
\dot p + \frac{p}{\tau_p} = M(1-\lambda)\bar\varepsilon_s\,\dot\xi,
\qquad
\tau_p = \frac{\eta_f\,\ell^{2}}{\kappa\,M},
\label{eq:hydraulic_relaxation}
\end{equation}
with $\kappa$ the Darcy permeability and $\eta_f$ the electrolyte
viscosity. Here $\ell$ denotes the drainage path length, which in the
lumped through-thickness model is the electrode thickness; the relevant
drainage geometry, and the resulting range of $\tau_p$, are reconsidered
in Sec.~\ref{sec: timescales}. In the frequency domain, using
Eq.~\ref{eq:porosity_response},
\begin{equation}
\hat p = \frac{M(1-\lambda)\bar\varepsilon_s\,\xi_0\,\beta}{1+i\omega\tau_\xi}\,
W(\omega)\,\hat c,
\qquad
W(\omega) = \frac{i\omega}{i\omega+\tau_p^{-1}} = \frac{i\omega\tau_p}{1+i\omega\tau_p},
\label{eq:pore_pressure_explicit}
\end{equation}
where $W(\omega)$ is the single-pole hydraulic relaxation factor: pore
pressure builds up under rapid porosity change ($W\to1$, undrained) and
vents to the separator at low frequency ($W\to i\omega\tau_p\to0$,
drained).

\paragraph{Poro-mechanical kernel}
The pore pressure enters the total stress through the Biot effective
stress relation $\hat\sigma_{\mathrm{tot}} = \hat\sigma^s - \alpha\hat p$,
so $-\alpha\hat p = \mathcal{H}_{pm}\hat c$ gives
\begin{equation}
\mathcal{H}_{pm}(\omega)
= -\frac{\alpha\,M(1-\lambda)\bar\varepsilon_s\,\xi_0\,\beta}{1+i\omega\tau_\xi}\,W(\omega)
= -\frac{\alpha\,M(1-\lambda)\bar\varepsilon_s\,\xi_0\,\beta\,i\omega}
        {(i\omega+\tau_p^{-1})(1+i\omega\tau_\xi)},
\label{eq:Hpm_specialized}
\end{equation}
with $\alpha$ the Biot coefficient.

\paragraph{The closed-form MEIS expression}
Inserting the kernels \eqref{eq:Hcm_specialized} and
\eqref{eq:Hpm_specialized} into the general structure
Eq.~\ref{eq:MEIS_general_structure}, with
$\mathcal{G}_e\mathcal{G}_c = t_+/(\mathrm{F}_{\!c}\,\ell\,i\omega)$, yields
the closed-form mechano-electrochemical impedance
\begin{equation}
Z_{\mathrm{MEIS}}(\omega)
= -\frac{t_+\,\beta}{\mathrm{F}_{\!c}\,\ell\,i\omega}
\left[\,E^*(\omega)\,\mathcal{B}(\omega)
+ \frac{\alpha\,M(1-\lambda)\bar\varepsilon_s\,\xi_0\,i\omega}
       {(i\omega+\tau_p^{-1})(1+i\omega\tau_\xi)}\,\right].
\label{eq:MEIS_closed_form}
\end{equation}
The prefactor $t_+\beta/(\mathrm{F}_{\!c}\,\ell\,i\omega)$ is the Faradaic
conversion of current into chemical eigenstrain; the first bracketed
term is the skeleton stress, modulated by viscoelastic relaxation
$E^*(\omega)$ and porosity accommodation $\mathcal{B}(\omega)$, and the
second is the pore-fluid pressurization, controlled by the Biot
coefficient $\alpha$, the void accommodation fraction $\lambda$, and the
coupled relaxation through $\tau_\xi$ and $\tau_p$.

The response is governed by four characteristic time scales: the
viscoelastic relaxation time $\tau_m$ (in $E^*$), the porosity
accommodation time $\tau_\xi$ (in both $\mathcal{B}$ and the pore-fluid
source), the hydraulic relaxation time $\tau_p$ (Darcy drainage), and the
chemical accumulation time $1/\omega$. Their relative ordering sets the
qualitative shape of the MEIS spectrum.

Three limiting cases connect Eq.~\ref{eq:MEIS_closed_form} to simpler
descriptions:
\begin{itemize}
\item \textbf{Rigid microstructure} ($k\to\infty$, $\xi_0\to0$):
porosity cannot rearrange, $\mathcal{B}\to1$, no fluid is displaced
($\hat p\to0$), and the impedance reduces to the chemo-viscoelastic
response $-t_+\beta E^*(\omega)/(\mathrm{F}_{\!c}\,\ell\,i\omega)$.
\item \textbf{Fully unsaturated} ($\lambda=1$): all porosity change
closes void space without displacing fluid; the pore-fluid term
vanishes and Eq.~\ref{eq:MEIS_closed_form} reduces to
$-t_+\beta E^*\mathcal{B}/(\mathrm{F}_{\!c}\,\ell\,i\omega)$, recovering
the bridge-function model of
Fang~et~al.~\citep{fang_mechano-electrochemical_2025}.
\item \textbf{Fully saturated} ($\lambda=0$): all porosity change
displaces fluid and the pore-fluid pressurization reaches its maximum;
porosity accommodation then reduces the skeleton-stress contribution
while enhancing the fluid-pressurization contribution, a trade-off
absent from either limit above.
\end{itemize}
Equation~\ref{eq:MEIS_closed_form} therefore unifies the phenomenological
bridge-function model and the classical saturated poro-mechanical model
within a single thermodynamically consistent framework, with $\lambda$
controlling the balance between kinematic accommodation and hydraulic
pressurization.

As in Section~\ref{sec: MEIS derivation}, Eq.~\ref{eq:MEIS_closed_form}
is written in the tension-positive convention; referred to the measured
compressive stack pressure $-\sigma_{\mathrm{tot}}$ it carries the
opposite, positive overall sign.

\section{Non-Dimensional Analysis of the MEIS Spectrum}
\label{sec: nondim}

\subsection{Dimensionless Form and Governing Parameters}
\label{sec: dimensionless form}

The closed-form impedance Eq.~\ref{eq:MEIS_closed_form} is governed by
four time scales---the viscoelastic relaxation $\tau_m$, the porosity
accommodation $\tau_\xi$, the hydraulic relaxation $\tau_p$, and the
inverse driving frequency $1/\omega$---together with the modulus
contrast of the skeleton and the strength of the poro-mechanical
coupling. To expose the independent groups that actually control the
spectrum, we non-dimensionalize. Scaling frequency by the viscoelastic
relaxation time,
\begin{equation}
\Omega = \omega\,\tau_m,
\end{equation}
the standard linear solid and the two relaxation factors become
\begin{equation}
\frac{E^*}{E_\infty} = \frac{1+i\lambda_E\Omega}{1+i\Omega},
\qquad
\mathcal{A}(\Omega) = \frac{1}{1+i\Omega/\Lambda_\xi},
\qquad
W(\Omega) = \frac{i\Omega}{i\Omega+\Lambda_p},
\end{equation}
where $\mathcal{A}$ is the porosity-accommodation factor and $W$ the
hydraulic-drainage factor. The dimensionless groups are
\begin{equation}
\lambda_E = \frac{E_0}{E_\infty}\ \ (\geq 1),
\qquad
\Lambda_\xi = \frac{\tau_m}{\tau_\xi},
\qquad
\Lambda_p = \frac{\tau_m}{\tau_p},
\qquad
\Pi = \frac{\alpha M(1-\lambda)\bar\varepsilon_s}{E_\infty},
\label{eq:dimensionless_groups}
\end{equation}
together with the equilibrium accommodation ratio $\xi_0$, which is
already dimensionless. Here $\lambda_E$ is the viscoelastic stiffness
contrast (instantaneous over relaxed modulus), $\Lambda_\xi$ and
$\Lambda_p$ measure how fast porosity rearrangement and pore-fluid
drainage are relative to skeleton relaxation, and $\Pi$ is the
poro-mechanical coupling strength---the ratio of the fluid storage
stiffness $\alpha M(1-\lambda)\bar\varepsilon_s$ to the skeleton
modulus $E_\infty$. The void accommodation fraction $\lambda$ and the
solid fraction $\bar\varepsilon_s$ therefore enter the spectrum only
through the single group $\Pi$. Normalizing the impedance by
\begin{equation}
Z_0 = \frac{t_+\,\beta\,E_\infty\,\tau_m}{\mathrm{F}_{\!c}\,\ell},
\label{eq:Z0}
\end{equation}
the dimensionless spectrum $\mathcal{Z}(\Omega)=Z_{\mathrm{MEIS}}/Z_0$ is
\begin{equation}
\mathcal{Z}(\Omega) =
-\,\frac{1+i\lambda_E\Omega}{i\Omega\,(1+i\Omega)}
\left(1 - \frac{\xi_0}{1+i\Omega/\Lambda_\xi}\right)
\;-\;
\frac{\Pi\,\xi_0}{(i\Omega+\Lambda_p)(1+i\Omega/\Lambda_\xi)}.
\label{eq:dimensionless_spectrum}
\end{equation}
The first term is the chemo-mechanical (skeleton) branch and the second
the poro-mechanical (pore-fluid) branch.

We treat the accommodation amplitude $\xi_0$ and time $\tau_\xi$ (hence
$\Lambda_\xi$) as constants, evaluated at a representative skeleton
modulus. Strictly, $\xi_0 = E^*/(E^*+k)$ and
$\tau_\xi = \eta_\xi/(E^*+k)$ inherit a weak frequency dependence
through $E^*(\omega)$; this is negligible when the microstructural
stiffness dominates the modulus contrast,
$k \gtrsim (E_0-E_\infty)$, and retaining it only renormalizes the
accommodation pole without changing the structure of
Eq.~\ref{eq:dimensionless_spectrum}.

\subsection{Structure of the Spectrum and Competition of Relaxation Mechanisms}
\label{sec: spectrum structure}

Three poles organize Eq.~\ref{eq:dimensionless_spectrum}: the
viscoelastic pole at $\Omega\sim 1$, the accommodation pole at
$\Omega\sim\Lambda_\xi$, and the hydraulic pole at $\Omega\sim\Lambda_p$,
all riding on the $1/(i\Omega)$ accumulation factor inherited from
chemical storage, as illustrated in Figure~\ref{fig:pole_analysis}.

\paragraph{Shared accommodation}
The decisive structural feature is that the accommodation factor
$\mathcal{A}(\Omega)$ enters \emph{both} branches. In the
chemo-mechanical branch it appears as $1-\xi_0\mathcal{A}$, reducing the
strain delivered to the skeleton: porosity rearrangement diverts part of
the chemical eigenstrain away from elastic loading. In the
poro-mechanical branch it appears as $+\xi_0\mathcal{A}$, since the same
porosity change is what displaces pore fluid and sources the pressure.
Increasing the accommodation ratio $\xi_0$ thus simultaneously
\emph{suppresses} the skeleton contribution and \emph{feeds} the
pore-fluid contribution; the two branches are coupled through this single
shared parameter rather than being independent. This is the
non-dimensional counterpart of the common branch point $\hat\xi$ in
Figs.~\ref{fig:chemo_mechanical} and~\ref{fig:poro_mechanical}.

\paragraph{Low-frequency limit ($\Omega\ll 1,\Lambda_\xi,\Lambda_p$)}
Here $\mathcal{A}\to1$, $E^*\to E_\infty$, and $W\to0$. The
chemo-mechanical branch reduces to $-(1-\xi_0)/(i\Omega)$, a
capacitive accumulation softened by the relaxed accommodation factor
$1-\xi_0=k/(E_\infty+k)$, while the poro-mechanical branch
approaches the constant offset $-\Pi\xi_0/\Lambda_p$. The mechanical
response is dominated by the diverging accumulation term, with the pore
fluid fully drained.

\paragraph{High-frequency limit ($\Omega\gg 1,\Lambda_\xi,\Lambda_p$)}
Here $\mathcal{A}\to0$, so the bridge $1-\xi_0\mathcal{A}\to1$ (porosity
frozen, full eigenstrain transmitted) and $E^*\to E_0$. The
chemo-mechanical branch tends to $-\lambda_E/(i\Omega)$, recovering the
stiff instantaneous skeleton, whereas the poro-mechanical branch decays
as $\Lambda_\xi/\Omega^{2}$ and vanishes: with the porosity frozen there
is no volume change to displace fluid, so the pore-fluid branch is
band-limited even though the drainage factor $W\to1$.

\paragraph{Competition}
The qualitative shape of the spectrum is set by the ordering of
$\Lambda_\xi$, $\Lambda_p$, and unity. When these are well separated,
the response splits into resolvable viscoelastic, accommodation, and
drainage features; when they are comparable, the corresponding
relaxations overlap and broaden in frequency. The poro-mechanical branch
is intrinsically band-limited, bounded below by drainage ($\Lambda_p$)
and above by frozen accommodation ($\Lambda_\xi$), and its weight is set
by $\Pi\xi_0$; the chemo-mechanical branch spans the whole range, its
low- and high-frequency plateaus separated by the factor
$\lambda_E/(1-\xi_0)$. Because the two branches add, MEIS resolves
chemical storage, skeleton viscoelasticity, porosity accommodation, and
pore-fluid drainage as distinct but coupled spectral signatures.

\definecolor{cblue}{rgb}{0.13,0.32,0.62}
\definecolor{cred}{rgb}{0.70,0.17,0.13}
\definecolor{cpur}{rgb}{0.45,0.20,0.60}
\newcommand{\polemark}[2]{%
\draw[#2,line width=1.1pt] ($(#1)+(-3.6pt,-3.6pt)$)--($(#1)+(3.6pt,3.6pt)$);
\draw[#2,line width=1.1pt] ($(#1)+(-3.6pt,3.6pt)$)--($(#1)+(3.6pt,-3.6pt)$);}
\newcommand{\zeromark}[2]{\draw[#2,line width=1.1pt,fill=white] (#1) circle (3.7pt);}
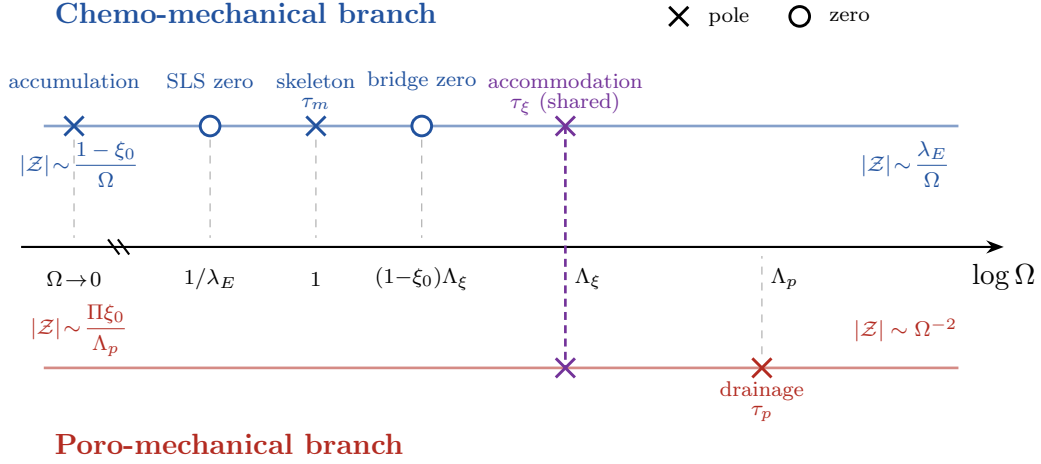
\begin{figure}[t]
\centering
\begin{tikzpicture}[font=\small,>=Stealth]
\def\yc{1.6}\def\yp{-1.6}
\draw[cblue!55,line width=1pt] (0.3,\yc)--(12.4,\yc);
\draw[cred!55,line width=1pt]  (0.3,\yp)--(12.4,\yp);
\draw[->,thick] (0,0)--(13.0,0) node[below=2pt]{$\log\Omega$};
\draw[thick] (1.15,0.10)--(1.30,-0.10); \draw[thick] (1.28,0.10)--(1.43,-0.10);

\node[cblue,font=\small\bfseries,anchor=west] at (0.3,\yc+1.5) {Chemo-mechanical branch};
\polemark{0.7,\yc}{cblue}
\node[cblue,font=\scriptsize,align=center] at (0.7,\yc+0.6) {accumulation};
\draw[dashed,black!35] (0.7,\yc-0.18)--(0.7,0.12);
\node[font=\scriptsize] at (0.7,-0.42) {$\Omega\!\to\!0$};
\zeromark{2.5,\yc}{cblue}
\node[cblue,font=\scriptsize,align=center] at (2.5,\yc+0.6) {SLS zero};
\draw[dashed,black!35] (2.5,\yc-0.18)--(2.5,0.12);
\node[font=\scriptsize] at (2.5,-0.42) {$1/\lambda_E$};
\polemark{3.9,\yc}{cblue}
\node[cblue,font=\scriptsize,align=center] at (3.9,\yc+0.6) {skeleton};
\node[cblue,font=\scriptsize,align=center] at (3.9,\yc+0.3) {$\tau_m$};
\draw[dashed,black!35] (3.9,\yc-0.18)--(3.9,0.12);
\node[font=\scriptsize] at (3.9,-0.42) {$1$};
\zeromark{5.3,\yc}{cblue}
\node[cblue,font=\scriptsize,align=center] at (5.3,\yc+0.6) {bridge zero};
\draw[dashed,black!35] (5.3,\yc-0.18)--(5.3,0.12);
\node[font=\scriptsize] at (5.3,-0.42) {$(1{-}\xi_0)\Lambda_\xi$};

\draw[cpur,line width=1pt,dash pattern=on 3pt off 2pt] (7.2,\yc)--(7.2,\yp);
\polemark{7.2,\yc}{cpur}
\polemark{7.2,\yp}{cpur}
\node[cpur,font=\scriptsize,align=center] at (7.2,\yc+0.6) {accommodation};
\node[cpur,font=\scriptsize,align=center] at (7.2,\yc+0.3) {$\tau_\xi$ (shared)};
\node[font=\scriptsize] at (7.5,-0.42) {$\Lambda_\xi$};

\node[cred,font=\small\bfseries,anchor=west] at (0.3,\yp-1.0) {Poro-mechanical branch};
\polemark{9.8,\yp}{cred}
\node[cred,font=\scriptsize,align=center] at (9.8,\yp-0.3) {drainage};
\node[cred,font=\scriptsize,align=center] at (9.8,\yp-0.6) {$\tau_p$};
\draw[dashed,black!35] (9.8,\yp+0.18)--(9.8,-0.12);
\node[font=\scriptsize] at (10.1,-0.42) {$\Lambda_p$};

\node[cblue,font=\scriptsize,align=center] at (0.77,\yc-0.50) {$|\mathcal{Z}|\!\sim\!\dfrac{1-\xi_0}{\Omega}$};
\node[cblue,font=\scriptsize,align=center] at (11.7,\yc-0.50) {$|\mathcal{Z}|\!\sim\!\dfrac{\lambda_E}{\Omega}$};
\node[cred,font=\scriptsize,align=center]  at (0.77,\yp+0.50) {$|\mathcal{Z}|\!\sim\!\dfrac{\Pi\xi_0}{\Lambda_p}$};
\node[cred,font=\scriptsize,align=center]  at (11.7,\yp+0.50) {$|\mathcal{Z}|\sim\Omega^{-2}$};

\polemark{8.7,3.05}{black}
\node[font=\scriptsize,anchor=west] at (8.95,3.05) {pole};
\zeromark{10.3,3.05}{black}
\node[font=\scriptsize,anchor=west] at (10.55,3.05) {zero};
\end{tikzpicture}
\caption{Pole--zero (corner-frequency) map of the dimensionless MEIS
spectrum, Eq.~\ref{eq:dimensionless_spectrum}, on the logarithmic
frequency axis $\Omega=\omega\tau_m$. The chemo-mechanical branch (top)
contains the accumulation pole at $\Omega\rightarrow0$, the
viscoelastic pole at $\Omega=1$ ($\tau_m$), the accommodation pole at
$\Omega=\Lambda_\xi$ ($\tau_\xi$), together with the viscoelastic zero
at $1/\lambda_E$ and the bridge zero at
$(1-\xi_0)\Lambda_\xi$. The poro-mechanical branch (bottom) contains the
drainage pole at $\Omega=\Lambda_p$ ($\tau_p$) and the shared
accommodation pole at $\Omega=\Lambda_\xi$ (purple). The common
accommodation pole reflects the shared porosity-accommodation mechanism
linking the chemo-mechanical and poro-mechanical branches. Feature
locations are schematic and depend on the relative values of
$\lambda_E$, $\Lambda_\xi$, and $\Lambda_p$.}
\label{fig:pole_analysis}
\end{figure}

\section{Results: Spectral Characteristics and Parametric Behavior}
\label{sec: results}

We now use the closed-form dimensionless spectrum
Eq.~\ref{eq:dimensionless_spectrum} to map how each governing parameter
shapes the mechano-electrochemical impedance. Throughout, spectra are
plotted in the measured (compression-positive) convention,
$\mathcal{Z}_{\rm meas}=-\mathcal{Z}$, so that the dominant capacitive
response occupies the first quadrant; frequency increases toward the
origin. We examine the Nyquist and Bode representations in turn, then
identify the conditions under which the spectrum crosses into the second
quadrant.

\subsection{Characteristic Frequencies}
\label{sec: characteristic frequencies}

The spectrum is organized by the poles and zeros catalogued in
Sec.~\ref{sec: spectrum structure} and mapped in
Fig.~\ref{fig:pole_analysis}. The chemo-mechanical branch carries the
accumulation pole at $\Omega\!\to\!0$, the viscoelastic pole at
$\Omega=1$, and the accommodation pole at $\Omega=\Lambda_\xi$, together
with the viscoelastic zero at $1/\lambda_E$ and the bridge zero at
$(1-\xi_0)\Lambda_\xi$; its magnitude rolls off as $\Omega^{-1}$ at both
ends, the two capacitive tails separated by the factor
$\lambda_E/(1-\xi_0)$. The poro-mechanical branch is a two-pole low-pass
sharing the accommodation pole $\Lambda_\xi$ and adding the drainage pole
$\Lambda_p$. Because the accommodation pole is common to both branches,
the porosity-accommodation parameters $\xi_0$ and $\Lambda_\xi$ couple
the skeleton and pore-fluid responses---a feature that recurs throughout
the parametric study below.

\subsection{Nyquist Representation}
\label{sec: nyquist results}

\begin{figure*}[h!]
\centering
\includegraphics[width=\textwidth]{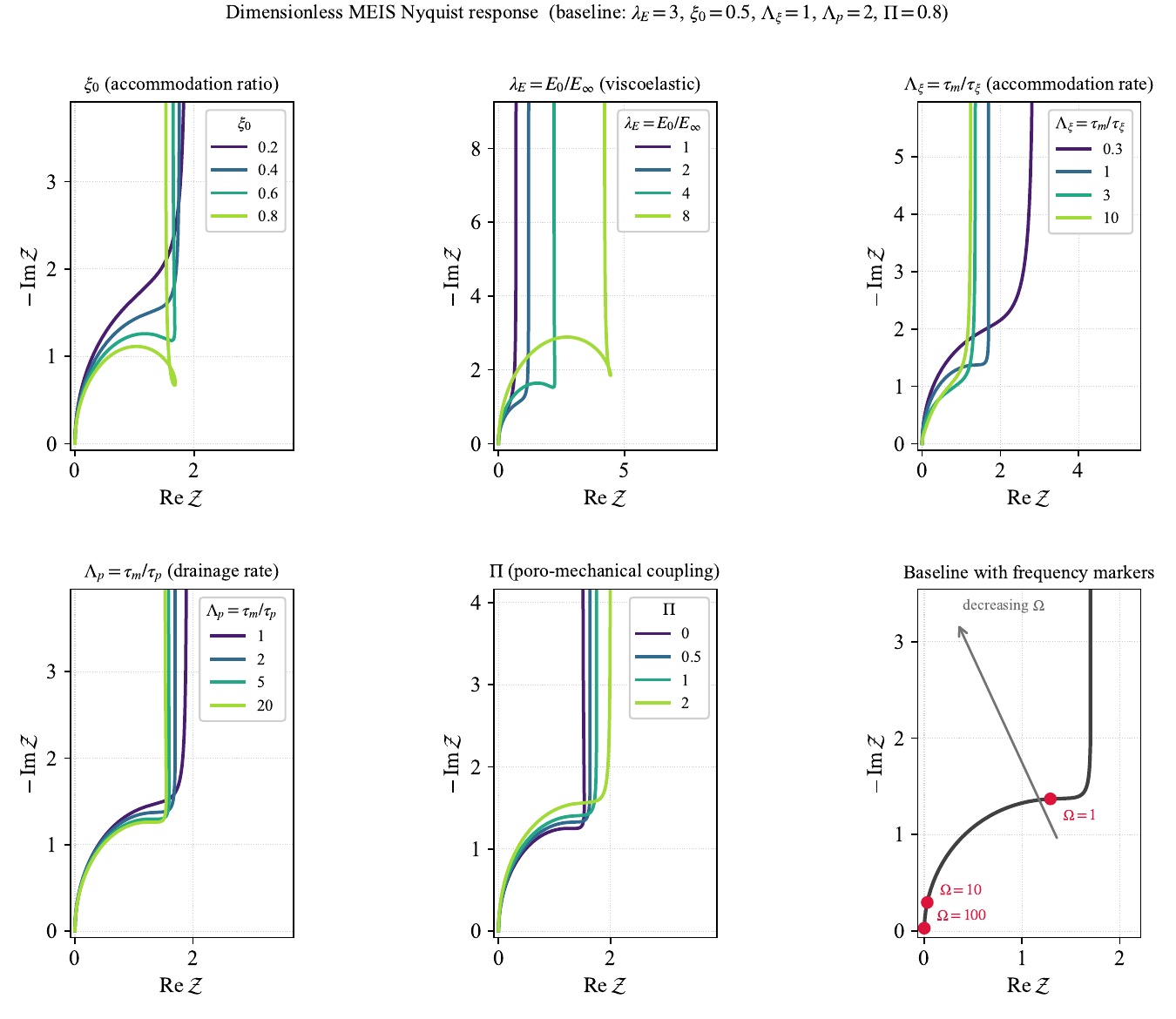}
\caption{Dimensionless MEIS Nyquist response
$\mathcal{Z}(\Omega)$ (Eq.~\ref{eq:dimensionless_spectrum}), plotted in
the measured (compression-positive) convention as
$-\mathrm{Im}\,\mathcal{Z}$ versus $\mathrm{Re}\,\mathcal{Z}$, with
frequency increasing toward the origin. Each panel varies one
dimensionless parameter while the others are held at the baseline
$\lambda_E=3$, $\xi_0=0.5$, $\Lambda_\xi=1$, $\Lambda_p=2$, and
$\Pi=0.8$. Panels (a)--(e) show the effects of the accommodation ratio
$\xi_0$, viscoelastic contrast $\lambda_E$, accommodation rate
$\Lambda_\xi$, drainage rate $\Lambda_p$, and poro-mechanical coupling
$\Pi$, respectively. Panel (f) shows the baseline response annotated
with representative frequency markers, illustrating the progression from
the high-frequency limit near the origin to the low-frequency
accumulation tail.}
\label{fig:nyquist_parametric}
\end{figure*}

Figure~\ref{fig:nyquist_parametric} sweeps each dimensionless group about
the baseline. Every spectrum shares the same qualitative form: a
depressed relaxation arc at intermediate frequency merging into a
near-vertical capacitive tail at low frequency. The parameters reshape
this locus in distinct ways. The accommodation ratio $\xi_0$ contracts
both the arc and the tail, since it lowers the real intercept
$\sim(1-\xi_0)(\lambda_E-1)+\xi_0/\Lambda_\xi$ and the tail weight
$(1-\xi_0)$; more accommodation diverts chemical strain from the skeleton
and softens the entire response. The viscoelastic contrast $\lambda_E$
opens the mid-frequency loop between the relaxed and instantaneous
skeleton stiffness, collapsing to a single capacitive line at
$\lambda_E=1$. The accommodation rate $\Lambda_\xi$ alters the arc
curvature as the porosity relaxation sweeps through the window. The
drainage rate $\Lambda_p$ produces only a small change, confirming that
the pore-fluid branch is a weak low-pass correction at moderate coupling.
The poro-mechanical coupling $\Pi$ adds real-axis weight at low and
intermediate frequency---the plateau $\sim\Pi\xi_0/\Lambda_p$---and
shifts the locus rightward, vanishing entirely at $\Pi=0$.

\subsection{Bode Representation}
\label{sec: bode results}

\begin{figure}[h!]
\centering
\includegraphics[width=\textwidth]{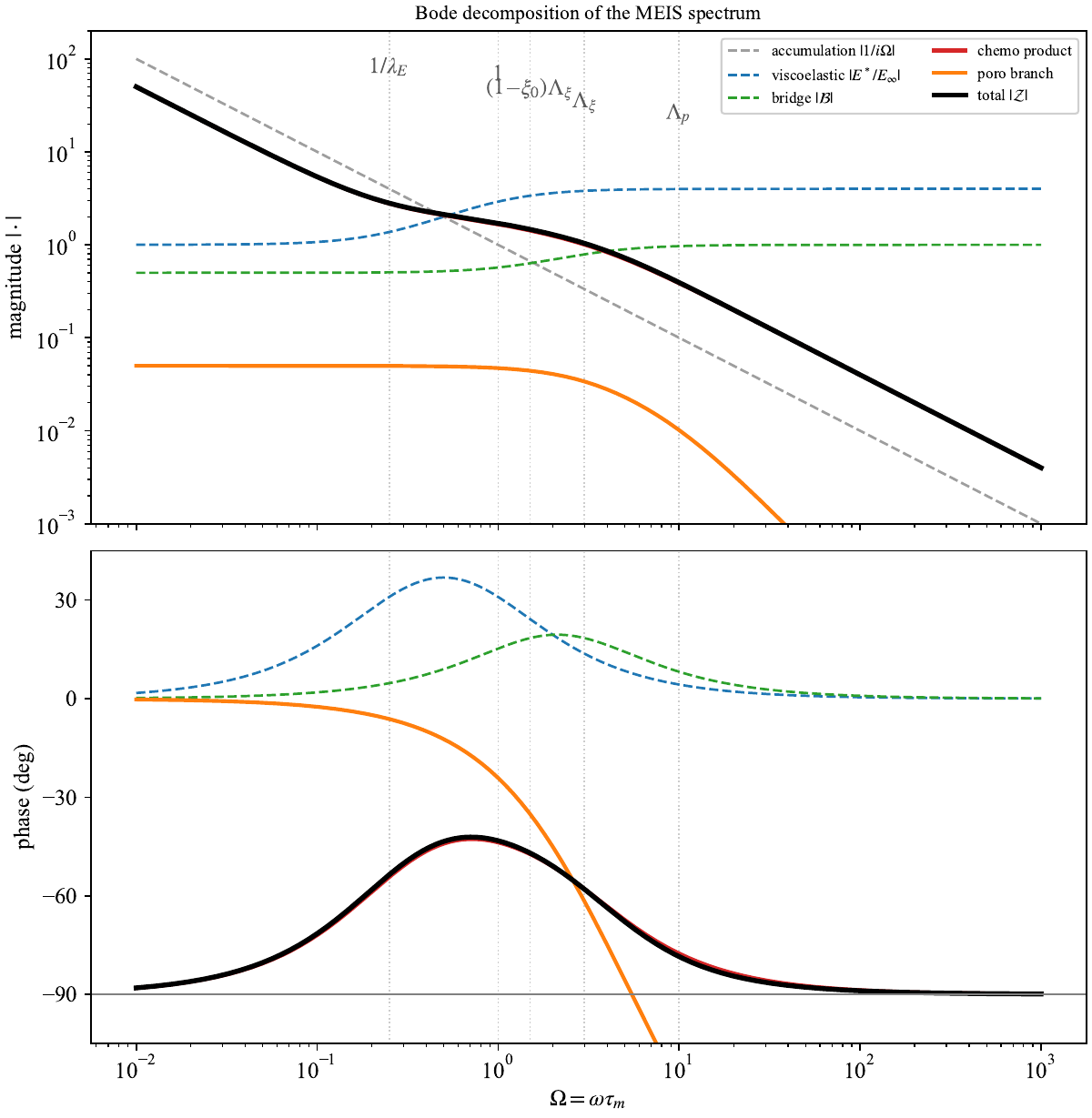}
\caption{Bode decomposition of the dimensionless MEIS spectrum in the
measured convention, with baseline parameters
$\lambda_E=4$, $\xi_0=0.5$, $\Lambda_\xi=3$, $\Lambda_p=10$, and
$\Pi=1$. The chemo-mechanical contribution is decomposed into the
multiplicative accumulation, viscoelastic, and bridge elements,
$\mathcal{Z}_{\rm chemo}=(1/i\Omega)(E^*/E_\infty)\mathcal{B}$, together
with their product, the poro-mechanical branch, and the total MEIS
response. The upper panel shows the magnitude and the lower panel the
phase as functions of the dimensionless frequency $\Omega$. Vertical
dotted lines indicate the characteristic corner frequencies
$1/\lambda_E$, $1$, $(1-\xi_0)\Lambda_\xi$, $\Lambda_\xi$, and
$\Lambda_p$ (cf. Fig.~\ref{fig:pole_analysis}).}
\label{fig:bode}
\end{figure}

The Nyquist locus is dominated by the $1/\Omega$ accumulation, which
compresses the relaxation features into the small near-origin region.
The Bode representation is more discriminating, because the
chemo-mechanical response is a \emph{product} of three multiplicative
elements,
\begin{equation}
\mathcal{Z}_{\rm chemo}
= \underbrace{\frac{1}{i\Omega}}_{\text{accumulation}}\;
  \underbrace{\frac{E^*}{E_\infty}}_{\text{viscoelastic}}\;
  \underbrace{\mathcal{B}}_{\text{bridge}},
\label{eq:bode_product}
\end{equation}
so that log-magnitudes and phases add (Fig.~\ref{fig:bode}). Removing the
accumulation factor exposes the mechanical relaxation directly:
$\Omega|\mathcal{Z}|$ steps from a low-frequency plateau equal to
$1-\xi_0$, the accommodation-softened modulus, to a high-frequency
plateau equal to $\lambda_E$, the stiff instantaneous modulus, while the
phase departs upward from the $-90^\circ$ capacitive floor by an amount
set by the same relaxations.

\begin{figure*}[h!]
\centering
\includegraphics[width=\textwidth]{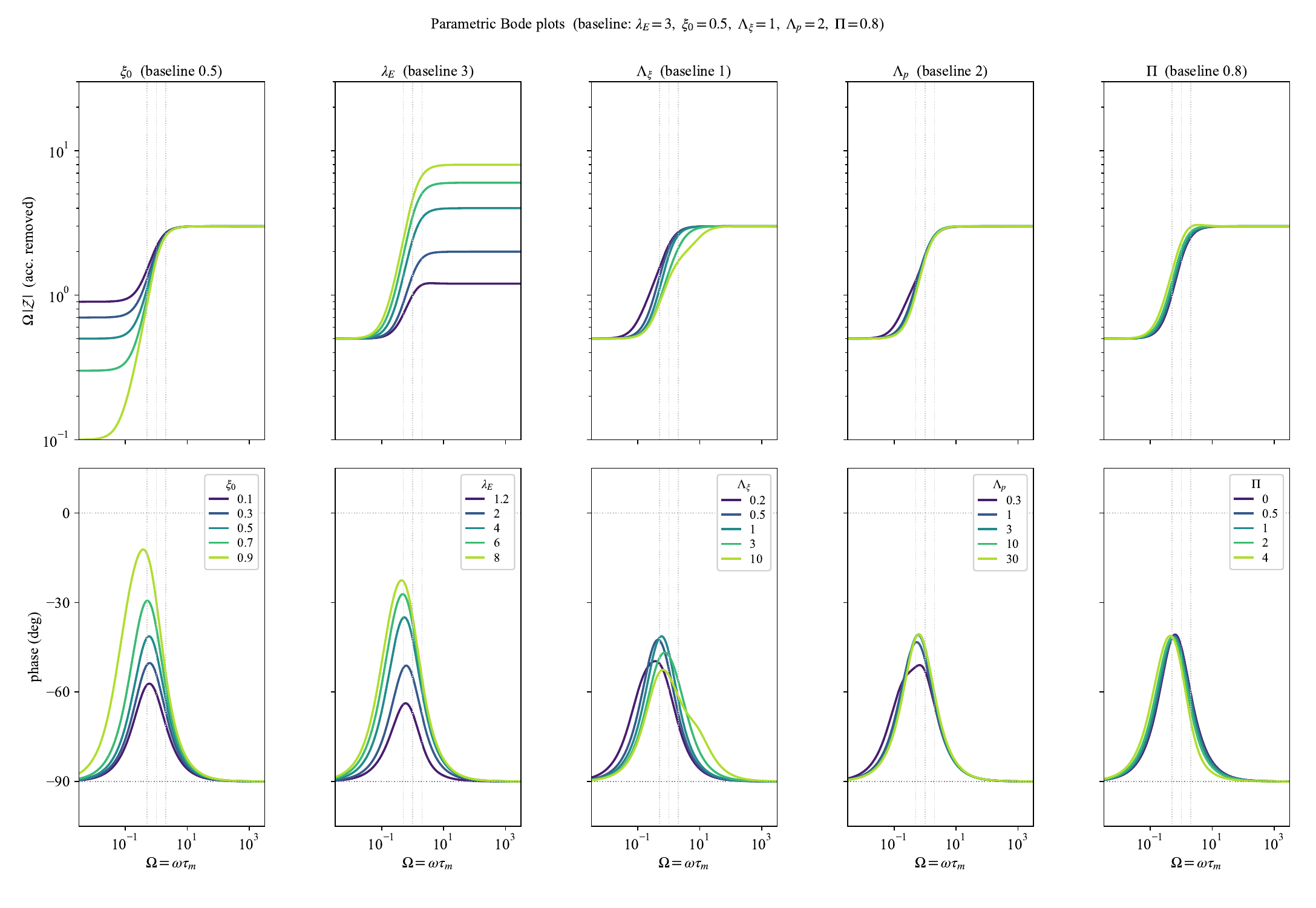}
\caption{Parametric Bode study of the dimensionless MEIS spectrum
(measured convention), varying one dimensionless parameter while the
others are held at the baseline $\lambda_E=3$, $\xi_0=0.5$,
$\Lambda_\xi=1$, $\Lambda_p=2$, and $\Pi=0.8$. The upper row shows the
accumulation-removed magnitude $\Omega|\mathcal{Z}|$, highlighting the
mechanical relaxation, while the lower row shows the corresponding phase.
Panels (a)--(e) illustrate the effects of the accommodation ratio
$\xi_0$, viscoelastic contrast $\lambda_E$, accommodation rate
$\Lambda_\xi$, drainage rate $\Lambda_p$, and poro-mechanical coupling
$\Pi$, respectively. Vertical dotted lines indicate the baseline corner
frequencies (cf. Fig.~\ref{fig:pole_analysis}).}
\label{fig:bode_param}
\end{figure*}

Figure~\ref{fig:bode_param} sweeps each parameter in this representation.
The accommodation ratio $\xi_0$ both lowers the low-frequency magnitude
plateau and \emph{deepens} the phase excursion, because increasing
$\xi_0$ separates the bridge zero $(1-\xi_0)\Lambda_\xi$ further below the
accommodation pole $\Lambda_\xi$; the peak phase rises from $-57^\circ$
at $\xi_0=0.1$ to $-12^\circ$ at $\xi_0=0.9$. The viscoelastic contrast
$\lambda_E$ sets the high-frequency magnitude plateau and similarly
enlarges the phase excursion (from $-64^\circ$ to $-22^\circ$ across
$\lambda_E=1.2$ to $8$), as the viscoelastic zero $1/\lambda_E$ separates
from its pole at $\Omega=1$. The accommodation rate $\Lambda_\xi$
translates the bridge feature horizontally along the frequency axis: the
phase excursion is maximal near $\Lambda_\xi\approx1$, where the bridge
feature reinforces the viscoelastic feature, and resolves into a separate
shoulder when $\Lambda_\xi$ departs from unity. In contrast, the
pore-fluid parameters are weak phase discriminators: the drainage rate
$\Lambda_p$ shifts the phase only for slow drainage
($\Lambda_p\lesssim1$, when the hydraulic pole enters the window), and
the coupling $\Pi$ shifts the phase-peak frequency slightly without
changing its height or the magnitude plateaus, since the pore-fluid
branch enters Eq.~\ref{eq:dimensionless_spectrum} additively rather than
multiplicatively.

This separation of roles has a direct practical consequence for fitting
MEIS data. The phase angle is governed almost entirely by the
chemo-mechanical parameters $(\xi_0,\lambda_E,\Lambda_\xi)$ and is nearly
insensitive to the pore-fluid parameters $(\Pi,\Lambda_p)$. The
accommodation and viscoelastic parameters are therefore best extracted
from the phase spectrum first, after which the pore-fluid coupling is
obtained from the residual low- and intermediate-frequency magnitude.

\subsection{Onset of Second-Quadrant Behavior}
\label{sec: q2 results}

\begin{figure*}[h!]
\centering
\includegraphics[width=\textwidth]{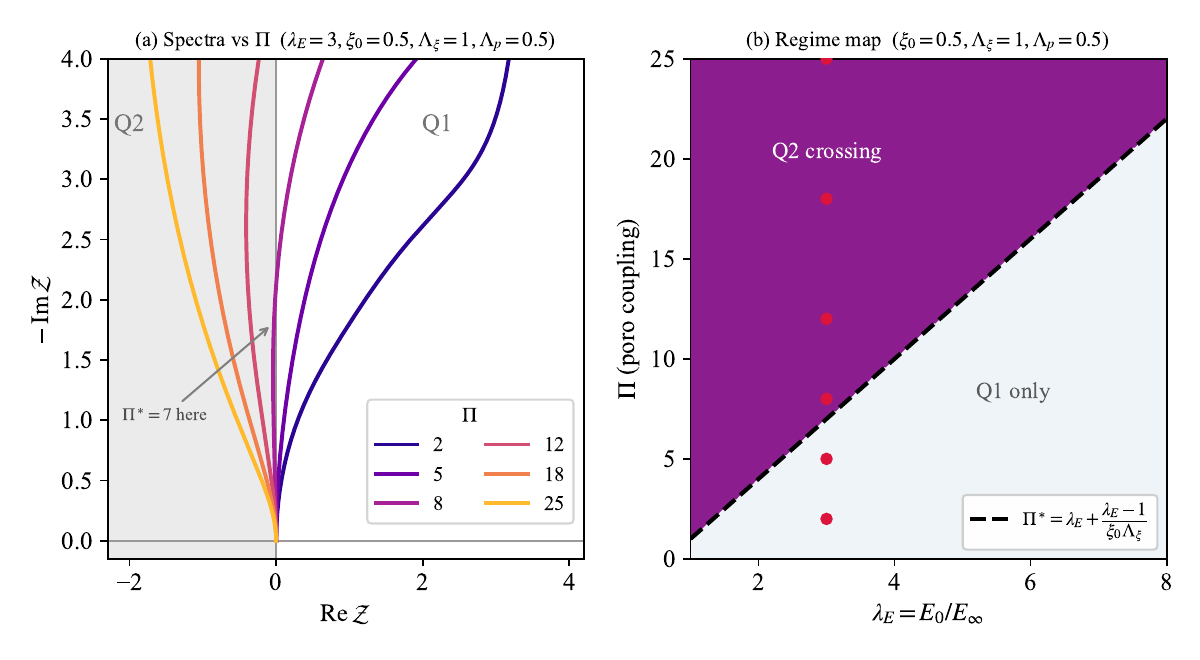}
\caption{Conditions for the MEIS spectrum to cross into the second
quadrant (negative in-phase response,
$\mathrm{Re}\,\mathcal{Z}<0$ in the measured convention). (a) Nyquist
spectra obtained by increasing the poro-mechanical coupling $\Pi$ while
holding $\lambda_E=3$, $\xi_0=0.5$, $\Lambda_\xi=1$, and
$\Lambda_p=0.5$ fixed. (b) Regime map in the
$(\lambda_E,\Pi)$ plane, showing regions where the spectrum remains
entirely in the first quadrant (Q1) and where second-quadrant (Q2)
crossing occurs. The dashed line denotes the analytical threshold
$\Pi^\ast=\lambda_E+(\lambda_E-1)/(\xi_0\Lambda_\xi)$
(Eq.~\ref{eq:Q2_threshold}), and the markers indicate the cases shown in
panel (a).}
\label{fig:second_quadrant}
\end{figure*}

A recurring experimental feature is the excursion of the spectrum into
the second quadrant ($\mathrm{Re}\,\mathcal{Z}<0$, a negative in-phase
response). Within the present single-electrode model this originates
solely in the poro-mechanical branch: the chemo-mechanical branch keeps a
phase lead (both of its zero--pole pairs place the zero below the pole)
and remains in the first quadrant, whereas the poro-mechanical branch is
a second-order low-pass whose real part turns negative above
$\Omega^\ast=\sqrt{\Lambda_p\Lambda_\xi}$. The spectrum crosses into the
second quadrant once this negative contribution overcomes the
chemo-mechanical branch, which occurs above the threshold
\begin{equation}
\Pi^\ast = \lambda_E + \frac{\lambda_E-1}{\xi_0\Lambda_\xi}
\;\xrightarrow{\ \xi_0\Lambda_\xi\gg1\ }\;\lambda_E,
\label{eq:Q2_threshold}
\end{equation}
i.e. when the pore-fluid storage stiffness exceeds the skeleton
stiffness, $\alpha M(1-\lambda)\bar\varepsilon_s\gtrsim E_0$.
Figure~\ref{fig:second_quadrant} shows the progression as $\Pi$ increases
through $\Pi^\ast$ and the corresponding regime map in the
$(\lambda_E,\Pi)$ plane: below $\Pi^\ast$ the spectrum is confined to the
first quadrant, while above it a loop near the origin
(intermediate-to-high frequency, deepest near $\Omega^\ast$) bends into
the second quadrant, the excursion growing with $\Pi/\lambda_E$ and with
slower drainage. The low-frequency tail, whose real intercept is strictly
positive, always remains in the first quadrant.

The single-electrode model therefore admits second-quadrant behavior only
under strong pore-fluid coupling and only at intermediate-to-high
frequency. The gradual, low-frequency second-quadrant entry frequently
seen in \emph{full-cell} measurements requires the competition between an
expanding and a contracting electrode, which we develop next in
Sec.~\ref{sec: full cell}.

\subsection{Full-Cell MEIS: Competition Between Electrodes}
\label{sec: full cell}

The single-electrode results above assume a rigid fixture
($\varepsilon=0$) acting on one active layer. A full-cell MEIS
measurement instead records the stack pressure of an
anode\,$|$\,separator\,$|$\,cathode assembly, in which a single applied
current produces \emph{anti-correlated} concentration changes in the two
electrodes---lithium leaving one enters the other---so that one electrode
expands while the other contracts. The measured stress is the result of
this competition, and it gives the full cell access to spectral regions
that no single electrode can reach.

\paragraph{Effective MEIS modulus}
Relaxing the rigid-fixture assumption, we retain the electrode strain
$\varepsilon$ in the porosity free energy, so the conjugate force becomes
$f_\xi=E^*(\beta c-\varepsilon)-(E^*+k)\xi$ and the kinetic law
$\dot\xi=f_\xi/\eta_\xi$ gives, in the frequency domain,
$\hat\xi=\Phi(\omega)(\beta\hat c-\hat\varepsilon)$ with
$\Phi=\xi_0/(1+i\omega\tau_\xi)$. Combining the skeleton stress
$\hat\sigma^s=E^*(\hat\varepsilon-\beta\hat c+\hat\xi)$, the pore pressure
$\hat p=M(1-\lambda)\bar\varepsilon_s W\hat\xi$, and the effective stress
$\hat\sigma=\hat\sigma^s-\alpha\hat p$, all terms collapse onto the single
relation
\begin{equation}
\hat\sigma = \hat E(\omega)\,(\hat\varepsilon-\beta\hat c),
\qquad
\hat E(\omega)=E^*-\Phi\big[E^*-\alpha M(1-\lambda)\bar\varepsilon_s W\big],
\label{eq:effective_modulus}
\end{equation}
where $\hat E(\omega)$ is the \emph{effective MEIS modulus}, carrying the
viscoelastic, accommodation, and pore-fluid relaxations in a single
complex stiffness. It reproduces the single-electrode result in the
blocked limit $\varepsilon=0$: since $-\hat E\beta=\mathcal{H}_{cm}+\mathcal{H}_{pm}$
(using $1-\Phi=\mathcal{B}$), the blocked impedance is
$Z=-\beta\hat E\,\Gamma$ with $\Gamma=t_+/(\mathrm{F}_{\!c}\ell\,i\omega)$
the accumulation factor, and $\hat E\to E_\infty(1-\xi_0)$ at low
frequency.

\paragraph{Series assembly under iso-stress}
We idealize the stack as three layers in mechanical series. Force balance
makes the normal stress common to all layers, and the rigid fixture fixes
the total thickness,
$\ell_a\hat\varepsilon_a+\ell_s\hat\varepsilon_s+\ell_c\hat\varepsilon_c=0$.
Each electrode obeys Eq.~\ref{eq:effective_modulus},
$\hat\varepsilon_k=\hat\sigma/\hat E_k+\beta_k\hat c_k$, while the
separator is elastic, $\hat\varepsilon_s=\hat\sigma/E_s$. Solving for the
common stress and inserting the anti-correlated sources
$\hat c_a=+\Gamma_a\hat J_e$, $\hat c_c=-\Gamma_c\hat J_e$ (the thickness
$\ell_k$ cancels against $\Gamma_k$) gives the full-cell impedance
\begin{equation}
Z_{\mathrm{MEIS}}^{\mathrm{cell}}(\omega)
= -\,\frac{\beta_a t_{+,a}-\beta_c t_{+,c}}
         {\mathrm{F}_{\!c}\,i\omega\,C_\Sigma(\omega)},
\qquad
C_\Sigma=\frac{\ell_a}{\hat E_a}+\frac{\ell_s}{E_s}+\frac{\ell_c}{\hat E_c},
\label{eq:Zcell}
\end{equation}
with $C_\Sigma$ the series (thickness-weighted) compliance. Equivalently,
in terms of the blocked half-cell impedances $Z_k=-\beta_k\hat E_k\Gamma_k$,
\begin{equation}
Z_{\mathrm{MEIS}}^{\mathrm{cell}} = w_a Z_a - w_c Z_c,
\qquad
w_k=\frac{\ell_k/\hat E_k}{C_\Sigma},
\qquad w_a+w_s+w_c=1,
\label{eq:weighted_difference}
\end{equation}
a compliance-weighted \emph{difference} of the two half-cell spectra; the
weights are complex and frequency dependent, so the softer electrode at a
given frequency dominates. Equation~\ref{eq:Zcell} reduces correctly to
the blocked anode impedance $Z_a$ for a rigid, inert cathode and rigid
separator.

\begin{figure*}[h!]
\centering
\includegraphics[width=\textwidth]{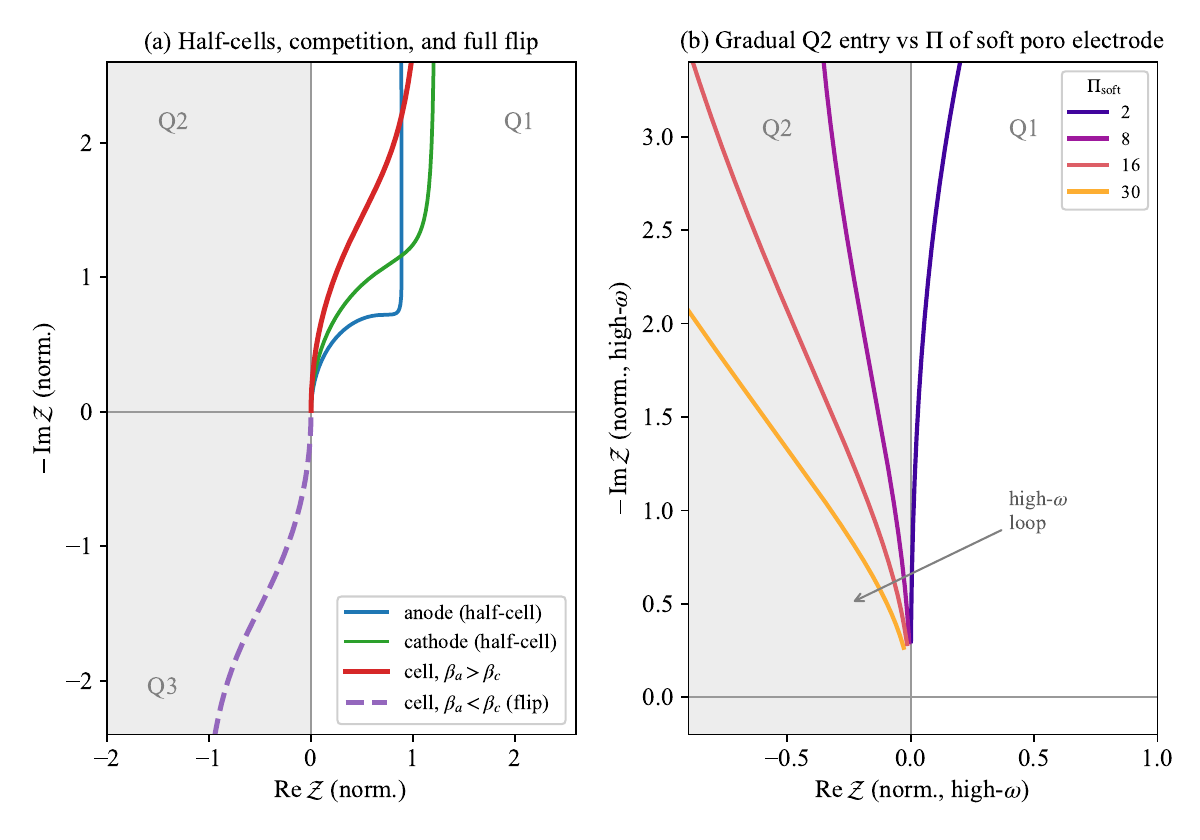}
\caption{Full-cell MEIS as a compliance-weighted difference of two
half-cells (Eqs.~\ref{eq:Zcell}--\ref{eq:weighted_difference}), measured
convention, normalized. (a) Two half-cell spectra (anode, cathode) lie in
the first quadrant; the cell combines them as a difference. When the
anode wins the chemical-strain competition ($\beta_a>\beta_c$) the cell
stays in Q1; when the cathode wins ($\beta_a<\beta_c$) the entire
spectrum reverses through the origin. (b) When one electrode is soft and
strongly pore-fluid--coupled, the cell bends gradually into the second
quadrant near the origin (high frequency), the excursion growing with
that electrode's coupling $\Pi_{\rm soft}$.}
\label{fig:fullcell}
\end{figure*}

\paragraph{Competition and second-quadrant behavior}
Writing $C_\Sigma=C_R+iC_I$ and $\Delta\equiv\beta_a t_{+,a}-\beta_c t_{+,c}$,
the measured response has the exact sign structure
\begin{equation}
\mathrm{Re}\,\mathcal{Z}_{\mathrm{meas}}
= \frac{\Delta}{\mathrm{F}_{\!c}}\,\frac{-C_I}{\omega|C_\Sigma|^2},
\qquad
-\mathrm{Im}\,\mathcal{Z}_{\mathrm{meas}}
= \frac{\Delta}{\mathrm{F}_{\!c}}\,\frac{C_R}{\omega|C_\Sigma|^2}.
\label{eq:cell_sign}
\end{equation}
Two mechanisms, neither available to a single electrode, move the cell
spectrum between quadrants. First, \emph{competition}: since $C_R>0$, the
sign of $\Delta$ selects the half-plane, so when the contracting
electrode wins the chemical-strain balance
($\beta_c t_{+,c}>\beta_a t_{+,a}$) the entire spectrum reverses through
the origin relative to the anode-dominated case; near electrode balance
$\Delta\to0$ and the mechanical signal collapses. Second, \emph{loss
reversal}: within a fixed half-plane the real part changes sign with
$C_I=-\sum_k\ell_k\,\mathrm{Im}(\hat E_k)/|\hat E_k|^2$. Two ordinary
(positive-loss) electrodes keep $C_I<0$ and remain in one quadrant, but a
soft, strongly pore-coupled electrode---whose effective loss modulus
turns negative above $\sqrt{\tau_p^{-1}\tau_\xi^{-1}}$, as in
Sec.~\ref{sec: q2 results}---can drive the compliance-weighted loss
negative over a band when it dominates $C_\Sigma$, bending the cell
gradually into the second quadrant there.
Figure~\ref{fig:fullcell} illustrates both: two half-cells confined to
the first quadrant combine into a cell that either reverses entirely with
the sign of $\Delta$ (the expand/contract competition) or develops a
partial second-quadrant loop when one electrode is soft and strongly
poro-coupled. The full cell thus reaches spectral regions inaccessible to
either electrode in isolation, providing a mechanistic origin for the
second-quadrant features observed in full-cell MEIS measurements.

\section{Discussion}
\label{sec:Discussion}

\subsection{Characteristic Time Scales and Parameter Estimates}
\label{sec: timescales}

The MEIS response is organized by three intrinsic relaxation times---the
viscoelastic time $\tau_m=\eta_m/E_1$ of the binder-supported skeleton,
the porosity-accommodation time $\tau_\xi=\eta_\xi/(E^*+k)$ of
microstructural rearrangement, and the hydraulic drainage time
$\tau_p=\eta_f L^2/(\kappa M)$ of the pore fluid---each probed against the
inverse driving frequency $1/\omega$. Whether a given relaxation is
resolvable depends on where it falls relative to the experimental window
(typically $\sim\!1$~mHz to $\sim\!10$~Hz, i.e. characteristic times
$\sim\!10^{-2}$ to $10^{2}$~s). We therefore estimate the three times and
the governing dimensionless groups from representative composite-electrode
parameters, collected in Table~\ref{tab:params}.

\begin{table}[t]
\centering
\caption{Representative physical parameters for a porous lithium-ion
composite electrode and the resulting MEIS time scales. Ranges are
order-of-magnitude literature estimates intended to bound the
dimensionless groups, not properties of a specific cell.}
\label{tab:params}
\small
\begin{tabular}{llll}
\toprule
Symbol & Quantity & Typical range & Note \\
\midrule
$E_\infty$        & relaxed electrode modulus      & $0.1$--$2$~GPa            & porosity/binder dependent \\
$\lambda_E=E_0/E_\infty$ & viscoelastic contrast   & $2$--$3$                  & modulus rises $\sim3\times$ on lithiation \\
$k$               & consolidation stiffness        & $0.1$--$2$~GPa            & resists porosity change \\
$\xi_0=E^*/(E^*+k)$ & accommodation ratio          & $0.3$--$0.7$              & shared by both branches \\
$\tau_m$          & viscoelastic relaxation        & $1$--$10^{4}$~s           & binder/polymer creep \\
$\tau_\xi$        & accommodation relaxation       & $1$--$10^{3}$~s           & particle rearrangement \\
$\eta_f$          & electrolyte viscosity          & $3$--$10$~mPa\,s          & carbonate electrolyte \\
$M$               & Biot modulus                   & $0.5$--$10$~GPa           & pore-fluid stiffness \\
$\kappa$          & permeability                   & $10^{-16}$--$10^{-14}$~m$^2$ & Kozeny--Carman \\
$\alpha$          & Biot coefficient               & $0.5$--$1$                & effective-stress coupling \\
$\lambda$         & void accommodation fraction    & $0$--$0.7$                & three-phase \\
$\bar\varepsilon_s$ & solid volume fraction        & $0.5$--$0.7$              & $1-\phi$ \\
$L$               & drainage length                & $10^{-6}$--$10^{-1}$~m    & particle $\to$ cell width \\
\midrule
$\tau_p$          & drainage relaxation            & $10^{-11}$--$10^{3}$~s    & $\propto L^{2}$ (see text) \\
\bottomrule
\end{tabular}
\end{table}

The viscoelastic and accommodation times both fall in the range
$1$--$10^{4}$~s, overlapping the experimental window: the binder creep and
the microstructural rearrangement are therefore directly resolvable in a
typical MEIS sweep, and the associated groups
$\lambda_E=E_0/E_\infty\sim2$--$3$ and $\xi_0\sim0.3$--$0.7$ are of order
unity. The poro-mechanical coupling
$\Pi=\alpha M(1-\lambda)\bar\varepsilon_s/E_\infty$ ranges from
$\sim\!0.02$ to $\sim\!70$ but is of order unity for representative
parameters; importantly, soft electrodes (small $E_\infty$) with stiff
pore fluid (large $M$) readily reach $\Pi\gtrsim\lambda_E$, the threshold
for second-quadrant behavior derived in Sec.~\ref{sec: q2 results}.

\begin{figure}[h!]
\centering
\includegraphics[width=\textwidth]{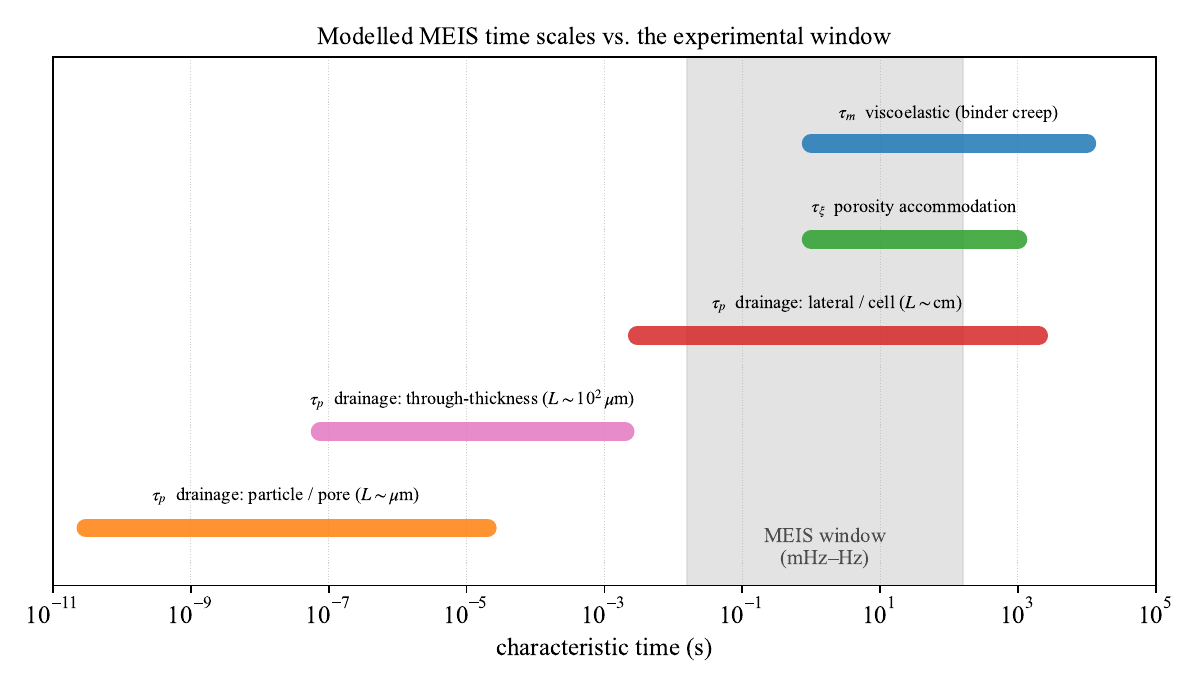}
\caption{Modelled MEIS relaxation times against the experimental window
(shaded, $\sim$mHz--Hz). The viscoelastic time $\tau_m$ and
accommodation time $\tau_\xi$ overlap the window and are resolvable. The
hydraulic time $\tau_p=\eta_f L^2/(\kappa M)$ depends strongly on the
drainage length $L$: particle/pore- and through-thickness drainage are
far faster than the window (the pore fluid is effectively undrained,
$W\to1$), whereas only lateral/cell-scale drainage produces a $\tau_p$
that enters the measurement window. The bars span the parameter ranges of
Table~\ref{tab:params}.}
\label{fig:timescales}
\end{figure}

\paragraph{The drainage length scale}
The hydraulic time is the outlier. Because $\tau_p\propto L^2$ and the
relevant Darcy length $L$ is not uniquely defined, $\tau_p$ spans roughly
fourteen orders of magnitude. Three candidate lengths give qualitatively
different pictures:
\begin{itemize}
\item \textbf{Particle/pore scale} ($L\sim1$--$10\,\mu$m):
$\tau_p\sim10^{-11}$--$10^{-5}$~s. Drainage is essentially instantaneous,
far faster than any accessible frequency.
\item \textbf{Through-thickness} ($L\sim50$--$100\,\mu$m):
$\tau_p\sim10^{-7}$--$10^{-3}$~s. Still below the window for all but the
slowest, lowest-permeability cases.
\item \textbf{Lateral / cell scale} ($L\sim1$--$10$~cm):
$\tau_p\sim10^{-3}$--$10^{3}$~s. Only here does $\tau_p$ enter the
experimental window.
\end{itemize}
Correspondingly the dimensionless drainage rate
$\Lambda_p=\tau_m/\tau_p$ spans some eighteen decades, from
$\sim\!10^{-3}$ (slow lateral drainage) to $\sim\!10^{14}$ (fast pore-scale
drainage), in contrast with $\Lambda_\xi=\tau_m/\tau_\xi$, which clusters
near unity. Figure~\ref{fig:timescales} places all of these against the
measurement window.

\paragraph{Implications for interpretation}
The wide span of $\Lambda_p$ has a concrete consequence. Pore-scale and
through-thickness drainage are far faster than the measurement
($\Lambda_p\gg1$, $W\to1$): the pore fluid is effectively undrained at all
observable frequencies and contributes only a frequency-independent
stiffening that is absorbed into the apparent modulus, leaving no
distinct hydraulic feature. A resolvable hydraulic relaxation in a MEIS
spectrum must therefore originate in \emph{long-range}, lateral or
edge-directed drainage across the electrode footprint, not in pore-scale
flow. This both identifies which $L$ to use when fitting $\tau_p$ and
cautions that an apparent ``poro-mechanical'' arc reflects cell-level
transport geometry rather than intrinsic microstructure. It also explains
why the drainage rate is a weak spectral discriminator at the
pore/thickness scale (Sec.~\ref{sec: bode results}): at those scales the
hydraulic pole simply lies outside the window.

\subsection{Limitations of the Linear Multiphysical Framework}
\label{sec: limitations}

The linear constitutive matrix developed in this work provides a
thermodynamically consistent and physically interpretable description of
the coupled electro--chemo--mechanical response of porous electrodes. It
is intended as a reduced spectral framework for identifying the dominant
relaxation mechanisms that shape the multiphysical impedance, rather than
a fully resolved microscopic account of every transport and deformation
process within the electrode. Several limitations follow from this scope.

First, the model is derived in the linear-response regime, in which the
constitutive coefficients are taken to be constant over the perturbation.
In practical electrodes the ionic mobility, mechanical stiffness, and
hydraulic permeability all vary appreciably with state of lithiation,
temperature, and microstructural evolution, so the extracted coefficients
should be read as effective quantities linearised about a reference
state, and the spectrum as the response of that state.

Second, transport and relaxation in porous electrodes are intrinsically
multiscale. Ionic motion may proceed within active particles, through the
electrolyte-filled pore network, or within interfacial double layers,
each with its own characteristic length and hence its own effective time;
viscoelastic deformation, pore-pressure equilibration, and microstructural
rearrangement are similarly distributed across scales. A measured MEIS
spectrum therefore superposes relaxation processes spread over broad
frequency bands rather than cleanly separated single-process responses.
The characteristic times $\tau_m$, $\tau_\xi$, and $\tau_p$ of the reduced
model (Sec.~\ref{sec: timescales}) should accordingly be understood as
effective spectral times marking dominant dynamical regimes, not uniquely
isolated microscopic constants.

Third, the formulation is homogenised at the electrode scale and does not
resolve particle-scale mechanics, evolving interparticle contact, or
microstructural fracture. The porosity-accommodation bridge function,
although derived here from a free-energy potential and therefore
thermodynamically consistent, represents this mesoscale structural
adaptation through effective parameters ($k$, $\eta_\xi$) rather than a
particle-resolved constitutive law.

Finally, the framework describes the bulk electrode response and does not
explicitly include interfacial charge-transfer kinetics. Quantitative
comparison with full-cell spectra---in particular the low-frequency
second-quadrant features discussed in Sec.~\ref{sec: full cell}---would
require coupling the present matrix with an appropriate electrochemical
interfacial model.

\subsection{Practical Application: Equivalent Rheological Networks}
\label{sec: application}

The closed-form derivation above is deliberately detailed, and the full
multiphysical expression may appear too involved for routine data
analysis. In practice it need not be inverted directly. Its essential
content is that the mechanical kernel of the MEIS response---the effective
modulus $\hat E(\omega)$---is a network of springs and dashpots. A
measured spectrum can therefore be fitted with a compact rheological
model in exactly the way an electrochemical impedance spectrum is fitted
with an equivalent electrical circuit, or a dynamic mechanical analysis
(DMA) curve with a generalized viscoelastic model. The physics-based
derivation then serves as the \emph{dictionary} that assigns a definite
physical meaning to each fitted element.

Each relaxation in the model corresponds to an identifiable mechanical
element (Figs.~\ref{fig:chemo_mechanical} and~\ref{fig:poro_mechanical}).
The viscoelastic skeleton is a standard linear solid---a spring $E_\infty$
in parallel with a Maxwell arm (spring $E_1$, dashpot $\eta_m$)---which
sets the stiffness contrast $\lambda_E=E_0/E_\infty$ and the time
$\tau_m=\eta_m/E_1$. Porosity accommodation is a Kelvin--Voigt element
(spring $k$, dashpot $\eta_\xi$) that diverts strain from the skeleton,
setting the accommodation ratio $\xi_0$ and the time $\tau_\xi$. Pore-fluid
drainage is a consolidation element (fluid storage stiffness $M$ in series
with a Darcy dashpot fixed by the permeability $\kappa$), setting the
coupling $\Pi$ and the time $\tau_p$. The chemical accumulation factor
$1/i\omega$, shared with the co-measured EIS, completes the network.

Because the chemo-mechanical elements combine multiplicatively
(Sec.~\ref{sec: bode results}), their signatures are separable in the
Bode representation, which makes the fit practical. The magnitude
plateaus fix the spring ratios---the low- and high-frequency, levels
$1-\xi_0$ and $\lambda_E$---while the phase bumps locate the dashpot time
constants $\tau_m$, $\tau_\xi$, and, where lateral drainage is resolved,
$\tau_p$. One therefore fits the phase to recover the chemo-mechanical
parameters and the residual magnitude to recover the pore-fluid coupling,
then maps the fitted spring constants and viscosities to $E_0$, $E_\infty$,
$k$, $\eta_m$, $\eta_\xi$, $M$, and $\kappa$ through the closed-form
relations of Eq.~\ref{eq:MEIS_closed_form}.

This is the same workflow long established for EIS and DMA, and it
inherits the same caveat: equivalent networks are not unique, and a
sufficiently elaborate network can reproduce any spectrum. The value of
the present derivation is that it fixes the minimal physically admissible
topology---how many springs and dashpots are warranted, and what each one
represents---so that the fit remains mechanistic rather than merely
descriptive. Furthermore, because MEIS and EIS are acquired synchronously
and share both the accumulation factor and several relaxation times, the
mechanical spring--dashpot network and the electrical equivalent circuit
can be fitted jointly, with common time constants constraining the two
spectra together. Tracking the fitted elements across state of charge or
cycle number then yields mechanistic diagnostics---stiffening,
accommodation loss, or altered drainage---each tied to a definite physical
quantity rather than to an empirical coefficient.

\subsection{A General Phenomenological MEIS Model for Fitting}
\label{sec: pheno fit}

The structure $Z_{\mathrm{MEIS}}=\mathcal{G}(\omega)\,\hat E(\omega)$,
with $\mathcal{G}\propto1/i\omega$ the chemical accumulation factor and
$\hat E(\omega)$ the effective mechanical modulus, suggests a general
form suitable for data fitting: any lumped network of springs and
dashpots has a modulus that is a rational function of $i\omega$, which
can always be written as a bank of elementary relaxation elements. We
therefore propose the phenomenological expression
\begin{equation}
Z_{\mathrm{MEIS}}(\omega)=\frac{K}{i\omega}\,\hat E(\omega),
\qquad
\hat E(\omega)=E_e
+\sum_{j=1}^{N} g_j\,\frac{i\omega\tau_j}{1+i\omega\tau_j}
+\sum_{p=1}^{P} h_p\,\frac{i\omega\tau_p^{\,d}}
                          {(1+i\omega\tau_p^{\,d})(1+i\omega\tau_p^{\,c})}.
\label{eq:pheno_general}
\end{equation}
Here $K$ is a real transduction gain (chemo-mechanical coupling per unit
current); $E_e$ is the equilibrium modulus; $\{g_j,\tau_j\}$ are the
strengths and times of $N$ \emph{viscoelastic} (Maxwell) relaxations
representing skeleton creep and porosity accommodation; and
$\{h_p,\tau_p^{\,d},\tau_p^{\,c}\}$ are the strengths and the two times
(drainage $\tau_p^{\,d}$, coupling $\tau_p^{\,c}$) of $P$
\emph{poromechanical} (consolidation) elements. The model is read off the
spectrum exactly as an equivalent circuit is fitted to EIS or a Prony
series to a DMA curve: one adds elements only until the residual is
explained, and each fitted element carries a definite mechanism.

\paragraph{Spectral content}
The modulus interpolates between $\hat E(0)=E_e$ (fully relaxed, fully
drained) and $\hat E(\infty)=E_e+\sum_j g_j$ (glassy), the two limits
separated by the relaxation strengths $g_j$. Writing $\hat E=E'+iE''$,
the impedance is $Z_{\mathrm{MEIS}}=(K/\omega)(E''-iE')$, so that
\begin{equation}
\mathrm{Re}\,Z_{\mathrm{MEIS}}\propto \frac{E''(\omega)}{\omega},
\qquad
\mathrm{Im}\,Z_{\mathrm{MEIS}}\propto -\frac{E'(\omega)}{\omega}:
\label{eq:Z_modulus_map}
\end{equation}
a MEIS Nyquist plot is, frequency by frequency, the complex mechanical
modulus---loss against storage, as in a Cole--Cole or DMA
representation---weighted by the chemical accumulation $1/\omega$. The
Maxwell terms contribute a strictly non-negative loss $E''$ and so remain
in the first quadrant. Each consolidation element, by contrast,
contributes a loss
$\propto h_p\,\omega\tau_p^{\,d}\,(1-\omega^2\tau_p^{\,d}\tau_p^{\,c})$
that changes sign at $\omega_p^\ast=(\tau_p^{\,d}\tau_p^{\,c})^{-1/2}$
while its storage part stays positive; when such an element dominates the
net loss above $\omega_p^\ast$, the spectrum crosses into the second
quadrant. The poromechanical family is thus the only one capable of
second-quadrant behavior, in agreement with Sec.~\ref{sec: q2 results}.

\paragraph{Relation to the physics-based model}
The closed-form result of Eq.~\ref{eq:MEIS_closed_form} is the minimal
instance $N=2,\,P=1$ of Eq.~\ref{eq:pheno_general}, which fixes the
parameter dictionary. The equilibrium modulus is $E_e=E_\infty(1-\xi_0)$;
the two viscoelastic times are the skeleton and accommodation times
$\tau_1=\tau_m$ and $\tau_2=\tau_\xi$, with strengths $g_1,g_2$ given by
the partial-fraction residues of $E^*(\omega)\mathcal{B}(\omega)$ (fixed
combinations of $E_\infty,E_0,\xi_0,\tau_m,\tau_\xi$); and the single
consolidation element has strength $h_1=\alpha M(1-\lambda)\bar\varepsilon_s\xi_0$
with drainage time $\tau_1^{\,d}=\tau_p$ and coupling time
$\tau_1^{\,c}=\tau_\xi$. The coupling time of the consolidation element
coincides with a viscoelastic time, reflecting the accommodation pole
shared by both branches of the physics model. Fitted parameters that do
not collapse onto this dictionary---additional relaxations, or
consolidation times that do not match an accommodation time---signal
physics beyond the present model (e.g. multiple drainage paths or
interfacial kinetics).

\paragraph{Practical fitting}
Equation~\ref{eq:pheno_general} inherits the strengths and the pitfalls
of equivalent-circuit analysis. The Bode separation of
Sec.~\ref{sec: bode results} makes the parameters identifiable: the
magnitude plateaus $E_e$ and $E_e+\sum_j g_j$ fix the spring strengths,
the phase peaks locate the times $\tau_j$, and a phase excursion past
$-90^\circ$ flags an active consolidation element and its
$\omega_p^\ast$. As in EIS, the representation is not unique and is prone
to over-fitting; the dictionary above provides the discipline, fixing how
many elements are physically admissible and what each represents. In
practice we recommend the smallest $N,P$ consistent with the residual,
fitted jointly with the co-measured EIS through their shared accumulation
factor and relaxation times.

\subsection{Toward Multiphysical Impedance Spectroscopy}
\label{sec: mpis}

The mechano-electrochemical impedance developed in this work is one
instance of a broader measurement principle. Its defining feature is not
that it is mechanical or electrochemical, but that it is a
\emph{cross-field} transfer function: a harmonic perturbation is applied
to a generalized force (or flux) of one physical field and the conjugate
flux (or force) of another field is recorded. In the current-controlled
configuration studied here the perturbation is an electrochemical flux,
the applied current, and the response is a mechanical force, the stack
stress, so the measured spectrum is precisely the transfer function
between the electrochemical and mechanical fields. Any system in which
several conjugate force--flux pairs are thermodynamically coupled admits
the same construction.

This is most transparent in the linear constitutive matrix of
Sec.~\ref{sec: General Framework}. Collecting the generalized fluxes---
electronic, ionic, solid deformation, pore fluid, and thermal---together
with their conjugate forces, the linear response is
$\mathbf{J}=\mathbf{L}\,\mathbf{X}$, in which the diagonal entries of
$\mathbf{L}$ are the direct, single-field responses and the off-diagonal
entries are the inter-field couplings (Table~\ref{tab:cross_effects}). A
conventional spectroscopy probes a diagonal block: electrochemical
impedance spectroscopy the electrical entry, dynamic mechanical analysis
the mechanical entry, thermal impedance the thermal entry. A
\emph{multiphysical} impedance spectroscopy instead probes an off-diagonal
block, perturbing field $a$ and measuring the conjugate variable of field
$b$, so that the transfer function $Z_{ba}(\omega)$ reduces in the static
limit to the coupling coefficient $L_{ba}$ and, across frequency, reports
the relaxation dynamics of that coupling.

The frequency dependence is the essential content. A static cross-field
measurement returns only the magnitude of a coupling; the spectrum
resolves the intermediate state variables---concentration, porosity, pore
pressure, temperature---through which the coupling is mediated, each
contributing its own characteristic time. For this reason the analysis
developed here is not specific to porous electrodes: the
non-dimensional groups, the characteristic-time hierarchy
(Sec.~\ref{sec: timescales}), and the equivalent spring--dashpot /
equivalent-circuit fitting (Secs.~\ref{sec: application}
and~\ref{sec: pheno fit}) all carry over to any coupled-field pair, with
only the identity of the elements and the parameter dictionary changing.

Several such spectroscopies follow immediately from the same matrix and
the couplings catalogued in Table~\ref{tab:cross_effects}. Perturbing
temperature and measuring current realizes a dynamic
thermo-electrochemical spectroscopy, the time-resolved Seebeck--Peltier
coupling; perturbing voltage and measuring fluid flux, or pressure and
measuring current, realizes electro-osmotic / streaming spectroscopy;
perturbing temperature and measuring stress or fluid flux realizes
thermo-mechanical (piezocaloric) or thermo-osmotic spectroscopy. MEIS
occupies the mechanical--electrochemical entry, while the remaining
entries define a family of largely unexplored coupled-field
spectroscopies. Moreover, because the constitutive matrix is symmetric
under the reciprocal (Onsager) relations, conjugate cross-spectroscopies
are not independent: $Z_{ba}$ and $Z_{ab}$ are linked, which provides an
internal consistency check and halves the number of measurements needed
to populate the coupling structure.

Seen in this light, mechano-electrochemical impedance spectroscopy is the
porous-electrode example of a general \emph{multiphysical impedance
spectroscopy}: apply a perturbation in one field, measure the response in
another, and read both the coupling and its dynamics from the resulting
transfer function. The thermodynamic framework, non-dimensionalization,
and fitting methodology established here for porous electrodes apply
wherever multiple physical fields are coupled, and we hope they provide a
common language for designing and interpreting such measurements across
electrochemistry, mechanics, and transport.

\section{Conclusions}
\label{sec: conclusions}

We have presented a thermodynamically consistent, closed-form theory of
mechano-electrochemical impedance spectroscopy (MEIS) for porous
electrodes. Beginning from a linear multiphysical constitutive matrix
that couples electronic, ionic, solid-deformation, pore-fluid, and
thermal fluxes to their conjugate forces, we reduced the coupled response
to a compact transfer function between the applied electrochemical
perturbation and the measured mechanical stress, making explicit the
physical content that a measured MEIS spectrum encodes.

Three elements distinguish the formulation. First, the
porosity-accommodation response, previously represented by a
phenomenological bridge function, was derived from a Helmholtz free
energy, so that the bridge function and its relaxation time follow from a
microstructural stiffness $k$ and viscosity $\eta_\xi$ rather than from a
fitted form. Second, a three-phase (solid--fluid--void) description with a
void-accommodation fraction $\lambda$ interpolates continuously between
the unsaturated and the fully saturated (Biot) limits, recovering earlier
two-phase models as special cases. Third, the impedance was organized
into a three-stage transfer-function structure---chemical accumulation
acting on the sum of a chemo-mechanical and a poro-mechanical branch---
yielding a single closed-form expression (Eq.~\ref{eq:MEIS_closed_form})
whose rigid, unsaturated, and saturated limits are each physically
transparent.

Non-dimensionalization reduced the response to five groups: the
viscoelastic contrast $\lambda_E$, the accommodation ratio $\xi_0$ and
rate $\Lambda_\xi$, the drainage rate $\Lambda_p$, and the
poro-mechanical coupling $\Pi$. A single accommodation pole was found to
couple the mechanical and pore-fluid branches. The parametric Nyquist and
Bode studies identified the phase angle as the sharp discriminator of the
chemo-mechanical parameters, while the pore-fluid coupling sets the
low-frequency magnitude---a separation that yields a concrete fitting
sequence. We further located the onset of second-quadrant behavior at
$\Pi^\ast=\lambda_E+(\lambda_E-1)/(\xi_0\Lambda_\xi)$ for a single
electrode, and showed that in a full cell the competition between an
expanding and a contracting electrode produces both a complete reversal
of the spectrum through the origin and the gradual second-quadrant entry
commonly observed in experiment---features inaccessible to either
electrode in isolation.

An estimate of the governing time scales placed the viscoelastic and
accommodation relaxations within the experimental window, but showed that
the hydraulic time, scaling with the square of an ambiguous drainage
length, spans many decades; only long-range, cell-scale drainage is
resolvable, so an observed poro-mechanical feature reflects transport
geometry rather than pore-scale structure. Despite the apparent
complexity of the full derivation, the mechanical kernel of the response
is a network of springs and dashpots. We therefore proposed a general
phenomenological expression (Eq.~\ref{eq:pheno_general})---a
generalized-Maxwell bank for the skeleton together with coupled
consolidation elements for the pore fluid---that can be fitted to data
exactly as equivalent circuits are fitted in EIS and Prony series in DMA.
The closed-form theory serves as the dictionary that assigns a physical
meaning to each fitted element and fixes the minimal admissible network,
so that the fit remains mechanistic rather than merely descriptive.

Finally, MEIS is only the porous-electrode instance of a general
principle: in any thermodynamically coupled system, perturbing one field
and recording the conjugate response of another yields a cross-field
transfer function---an off-diagonal entry of the constitutive matrix---
whose frequency dependence carries the dynamics of the coupling. The same
framework, non-dimensionalization, and equivalent-network fitting extend
to thermo-electrochemical, electro-osmotic, thermo-mechanical, and other
multiphysical impedance spectroscopies, with the reciprocal relations
linking conjugate measurements. Promising directions include quantitative
validation against full-cell MEIS data, coupling the present bulk
framework to interfacial charge-transfer kinetics to capture the
lowest-frequency response, and exploration of the off-diagonal
spectroscopies that this framework brings within reach. It is also worth noting that the linear response function for MEIS is important not only for the frequency domain measurements, but also as the general transfer function for linear response in the time domain (Laplace transform) and for steady state couplings.

\section*{Acknowledgment}
This work was supported by the National Science Foundation through award
CMMI-2543158 (CAREER program). We also gratefully acknowledge the Toyota
Research Institute of North America (TRINA), whose Electrochemical Society
Toyota Young Investigator Fellowship supported this research at its
earliest stage, when it was still an unproven idea; that early gift
allowed the concept to mature into the present NSF CAREER project. We
thank Professor Martin Z. Bazant of MIT for acting as an unofficial
advisor to the team, and for guidance that helped us develop
multiphysical impedance spectroscopy within a rigorous mathematical
framework grounded in irreversible thermodynamics.

\section*{Author contributions: CRediT}
\textbf{Junning Jiao:} Data curation; Formal analysis; Investigation; Methodology; Software; Validation; Visualization; Writing – original draft. 
\textbf{Juner Zhu:} Conceptualization; Data curation; Formal analysis; Investigation; Methodology; Project administration; Resources; Validation; Visualization; Supervision; Writing – original draft; Writing – review \& editing; Funding acquisition.

\appendix
\section{Complete Derivation of the MEIS Transfer Function}
\label{app: full derivation}

This appendix presents the full step-by-step derivation of the MEIS
transfer function, retaining all terms appearing in the reduced Onsager
structure without dropping any couplings or relaxation contributions.
All fields are treated under a lumped (thickness-averaged)
approximation, in which through-thickness gradients are represented by
finite differences across the electrode. The result is written in
closed form, with subsequent sections of the main text showing how this
general expression reduces to the standard MEIS formula under common
simplifying assumptions.

\subsection{Starting Point: Constitutive Relations and Conservation Laws}

We begin with the four constitutive relations from the reduced
constitutive matrix structure of Section~\ref{sec: Reduced Onsager Structure},
written along the through-thickness direction $x \in [0, \ell]$:
\begin{align}
J_e        &= L_{11}(-\partial_x\phi) + L_{12}(-\partial_x\mu),
              \label{eq:appC1}\\
F_i        &= L_{12}(-\partial_x\phi) + L_{22}(-\partial_x\mu)
              + L_{23}(-\sigma^p), \label{eq:appC2}\\
\dot{\varepsilon}^p &= L_{23}(-\partial_x\mu) + L_{33}(-\sigma^p)
              + L_{34}(-\partial_x p), \label{eq:appC3}\\
Q          &= L_{34}(-\sigma^p) + L_{44}(-\partial_x p).
              \label{eq:appC4}
\end{align}
Onsager reciprocity ($L_{ij} = L_{ji}$) has been imposed throughout.

The four conservation laws are
\begin{align}
\partial_x J_e &= 0, \label{eq:appB1}\\
\partial_t c + \partial_x F_i &= 0, \label{eq:appB2}\\
\partial_x \sigma^p &= 0, \label{eq:appB3}\\
\partial_t \zeta + \partial_x Q &= 0, \label{eq:appB4}
\end{align}
where $\zeta$ is the increment of fluid content (fluid volume gained per
unit reference volume), the natural Biot storage variable.

The boundary conditions for a single electrode in the stack are:
\begin{itemize}
    \item $J_e(0,t) = J_e(\ell, t) = \hat{J}_e e^{i\omega t}$
    (applied current);
    \item $F_i(0, t) = (t_+/\mathrm{F}_{\!c})J_e(t)$ (Faradaic flux at
    the separator), $F_i(\ell, t) = 0$ (blocking at the impermeable
    current collector);
    \item $\hat p(0,t) = 0$ (drained: the separator is hydraulically
    connected to the electrolyte reservoir), $\hat Q(\ell,t) = 0$
    (sealed: impermeable current collector);
    \item Global mechanical constraint: $\varepsilon(t) = 0$
    (single-electrode lumped form).
\end{itemize}

\subsection{Frequency-Domain Transformation}

Under harmonic excitation $\hat{J}_e e^{i\omega t}$, all fields
oscillate at the same frequency, so time derivatives transform
according to $\partial_t \to i\omega$. The conservation laws become
\begin{align}
\partial_x \hat{J}_e &= 0 \quad \Rightarrow \quad
\hat{J}_e(x,\omega) = \hat{J}_e(\omega) \text{ uniform}, \\
i\omega\,\hat{c} + \partial_x \hat{F}_i &= 0, \\
\partial_x \hat{\sigma}^p &= 0 \quad \Rightarrow \quad
\hat{\sigma}^p(x,\omega) = \hat{\sigma}^p(\omega) \text{ uniform}, \\
i\omega\,\hat{\zeta} + \partial_x \hat{Q} &= 0.
\end{align}

\subsection{Lumped Spatial Approximation}

All fields are treated under a lumped approximation, in which gradients
are represented by through-thickness finite differences,
$\partial_x \hat{f}(x,\omega) \approx \ell^{-1}\hat{f}(\omega)\big|_0^\ell$,
and the transport balances are closed by integrating each conservation
law across the thickness and applying the boundary fluxes.
\begin{itemize}
    \item \textbf{Mechanical.} The global constraint
    $\int_0^\ell \dot{\varepsilon}^p\,dx = 0$ reduces to the pointwise
    condition $\hat{\dot{\varepsilon}}^p = 0$.
    \item \textbf{Electrochemical.} Integrating the species
    conservation law $i\omega\hat{c} + \partial_x\hat{F}_i = 0$ across
    the thickness and applying the blocking condition
    $\hat{F}_i(\ell) = 0$ at the current collector gives
    $i\omega\ell\,\hat c = \hat F_i(0)-\hat F_i(\ell) = \hat F_i(0)$:
    the rate of species accumulation equals the ionic flux entering at
    the separator face.
    \item \textbf{Hydraulic.} Identically, integrating
    $i\omega\hat\zeta + \partial_x\hat Q = 0$ across the thickness and
    applying the sealed-collector condition $\hat Q(\ell)=0$ gives
    $i\omega\ell\,\hat\zeta = \hat Q(0)$: the rate of fluid-content
    change equals the drainage flux at the drained separator. With the
    lumped Darcy flux $\hat Q(0)\approx -(L_{44}/\ell)\hat p$, the
    single representative pore pressure $\hat p$ relaxes on the
    hydraulic time $\tau_p$, as derived in Step~4.
\end{itemize}

\subsection{Step 1: Elimination of the Electric Potential}

Solving Eq.~\ref{eq:appC1} for $-\partial_x\hat\phi$:
\begin{equation}
-\partial_x\hat\phi = \frac{\hat{J}_e}{L_{11}}
                    - \frac{L_{12}}{L_{11}}(-\partial_x\hat\mu).
\label{eq:appE1}
\end{equation}
Substituting into Eq.~\ref{eq:appC2}:
\begin{equation}
\hat{F}_i = L_{12}\left[\frac{\hat{J}_e}{L_{11}}
            - \frac{L_{12}}{L_{11}}(-\partial_x\hat\mu)\right]
            + L_{22}(-\partial_x\hat\mu) + L_{23}(-\hat\sigma^p).
\end{equation}
Combining the $\partial_x\hat\mu$ terms:
\begin{equation}
\hat{F}_i = \frac{L_{12}}{L_{11}}\hat{J}_e
          + \widetilde{L}_{22}(-\partial_x\hat\mu)
          + L_{23}(-\hat\sigma^p),
\label{eq:appFi_reduced}
\end{equation}
where
\begin{equation}
\widetilde{L}_{22} \equiv L_{22} - \frac{L_{12}^{\,2}}{L_{11}}
\label{eq:app_Ltilde}
\end{equation}
is the effective ionic transport coefficient after eliminating the
electronic degree of freedom.

\subsubsection*{Identification of the cation transference number.}
In the absence of all driving forces other than the imposed current
($\partial_x\hat\mu = 0$, $\hat\sigma^p = 0$), Eq.~\ref{eq:appFi_reduced}
reduces to $\hat{F}_i = (L_{12}/L_{11})\hat{J}_e$. The cation current
$\mathrm{F}_{\!c}\hat{F}_i$ is then a fraction $\mathrm{F}_{\!c}L_{12}/L_{11}$
of the total electronic current. This fraction is, by definition,
the cation transference number:
\begin{equation}
t_+ \equiv \mathrm{F}_{\!c}\,\frac{L_{12}}{L_{11}}.
\label{eq:app_transference}
\end{equation}

\subsection{Step 2: Closing the Ionic Sub-Problem}

Under the lumped approximation, the ionic balance is represented by $i\omega\ell\,\hat c = \hat F_i$,
which equates the rate of species accumulation to a representative
ionic flux through the electrode thickness. Evaluating $\hat F_i$
from the constitutive relation Eq.~\ref{eq:appFi_reduced}, with the
lumped gradient $-\partial_x\hat\mu \approx -\hat\mu/\ell$,
\begin{equation}
i\omega\ell\,\hat{c} = \frac{L_{12}}{L_{11}}\hat{J}_e
                    - \frac{\widetilde{L}_{22}}{\ell}\hat\mu
                    + L_{23}(-\hat\sigma^p).
\label{eq:app_ionic_balance}
\end{equation}
In the limiting case of no concentration gradient and no mechanical
driving, this reduces to $i\omega\ell\hat c = (L_{12}/L_{11})\hat J_e
= (t_+/\mathrm{F}_{\!c})\hat J_e$, consistent with the Faradaic boundary
condition and with the definition of $t_+$ in
Eq.~\ref{eq:app_transference}.

At this stage, $\hat\mu$ has not been related to $\hat{c}$. We
introduce the thermodynamic susceptibility
\begin{equation}
\chi \equiv \left.\frac{\partial\mu}{\partial c}\right|_{\text{ref}},
\label{eq:app_chi_def}
\end{equation}
evaluated at the reference state. For an ideal dilute solution,
$\chi = \mathrm{R}T/\bar{c}$; for non-ideal solutions, $\chi$ includes
activity corrections. Under the linearization $\hat\mu = \chi\hat{c}$
in the lumped approximation:
\begin{equation}
i\omega\ell\,\hat{c} = \frac{L_{12}}{L_{11}}\hat{J}_e
                    - \frac{\widetilde{L}_{22}\chi}{\ell}\hat{c}
                    - L_{23}\hat\sigma^p.
\end{equation}
Rearranging to collect $\hat{c}$ on the left:
\begin{equation}
\left(i\omega\ell + \frac{\widetilde{L}_{22}\chi}{\ell}\right)\hat{c}
= \frac{L_{12}}{L_{11}}\hat{J}_e - L_{23}\hat\sigma^p.
\end{equation}
Dividing by $\ell$ and introducing the diffusion time scale
\begin{equation}
\tau_D \equiv \frac{\ell^2}{\widetilde{L}_{22}\chi},
\label{eq:app_tauD_def}
\end{equation}
we obtain
\begin{equation}
\left(i\omega + \tau_D^{-1}\right)\hat{c}
= \frac{1}{\ell}\left(\frac{L_{12}}{L_{11}}\hat{J}_e
                      - L_{23}\hat\sigma^p\right).
\label{eq:app_concentration_implicit}
\end{equation}
This is the \textbf{first key intermediate result}: the concentration
amplitude is determined by both the applied current and the
chemo-mechanical feedback through $L_{23}\hat\sigma^p$.

\subsection{Step 3: Mechanical Constraint and Skeleton Stress}

Under the lumped approximation, the constraint $\hat{\dot\varepsilon}^p = 0$
applied to Eq.~\ref{eq:appC3} (with $-\partial_x\hat\mu \approx
-\hat\mu/\ell = -\chi\hat{c}/\ell$ and $-\partial_x\hat{p} \approx
-\hat{p}/\ell$) gives:
\begin{equation}
0 = L_{23}\left(-\frac{\chi\hat{c}}{\ell}\right)
   + L_{33}(-\hat\sigma^p)
   + L_{34}\left(-\frac{\hat{p}}{\ell}\right),
\end{equation}
or
\begin{equation}
L_{33}\hat\sigma^p = -\frac{L_{23}\chi}{\ell}\hat{c}
                   - \frac{L_{34}}{\ell}\hat{p}.
\label{eq:app_stress_balance}
\end{equation}
Solving for $\hat\sigma^p$:
\begin{equation}
\hat\sigma^p = -\frac{L_{23}\chi}{L_{33}\ell}\hat{c}
             - \frac{L_{34}}{L_{33}\ell}\hat{p}.
\label{eq:app_sigma_p_explicit}
\end{equation}
Here $\hat\sigma^p$ is the \emph{porous-medium stress} from the Onsager
constitutive law, corresponding to the dissipative mechanical response
governed by $L_{33}$. To obtain the total load-cell stress we combine
this with the pore-pressure contribution via the Biot effective stress
relation; this is handled in Step~5.

\subsection{Step 4: Pore-Pressure Closure}

The pore-fluid flux from Eq.~\ref{eq:appC4} is
\begin{equation}
\hat{Q} = L_{34}(-\hat\sigma^p) + L_{44}(-\partial_x\hat p).
\end{equation}
Because mechanical equilibrium (Eq.~\ref{eq:appB3}) renders
$\hat\sigma^p$ spatially uniform, the $L_{34}\hat\sigma^p$ contribution
is divergence-free and does not drive fluid redistribution; only the
Darcy term $L_{44}(-\partial_x\hat p)$ generates flow. Under the lumped
approximation the pressure gradient is represented by the
through-thickness finite difference $-\partial_x\hat p \approx
-\hat p/\ell$, so the Darcy flux is $\hat Q \approx -(L_{44}/\ell)\hat p$,
draining from the sealed collector ($\hat Q(\ell)=0$) toward the drained
separator ($\hat p(0)=0$).

\paragraph{Storage relation}
Here $\hat\xi$ is the relative pore \emph{contraction}, an internal
kinematic variable measuring the portion of particle-scale chemical
expansion absorbed by closure of the pore space (its strain
decomposition and constitutive evolution are given in Step~5). With
$\lambda\in[0,1]$ the void accommodation fraction, and a fraction
$(1-\lambda)$ of the porosity change displacing fluid, the fluid-content
increment is
\begin{equation}
\hat\zeta = -(1-\lambda)\,\bar\varepsilon_s\,\hat\xi + \frac{\hat p}{M},
\label{eq:app_storage_corrected}
\end{equation}
where $\bar\varepsilon_s$ is the steady-state solid volume fraction and
$M$ the Biot modulus. The leading minus sign is fixed by physical
consistency: a positive pore contraction $\hat\xi>0$ must \emph{expel}
fluid, so that in the undrained limit ($\hat\zeta=0$) the pore pressure
\emph{rises}, $\hat p_{\mathrm u} = M(1-\lambda)\bar\varepsilon_s\hat\xi
>0$. Writing the storage in terms of the fluid-content increment
$\zeta$ rather than the fluid volume fraction removes the sign
ambiguity that otherwise arises from conflating fluid storage with
pore-space kinematics.

\paragraph{Lumped hydraulic balance}
Integrating the fluid continuity equation
$i\omega\hat\zeta + \partial_x\hat Q = 0$ across the thickness and
applying the boundary fluxes gives
\begin{equation}
i\omega\ell\,\hat\zeta = \hat Q(0)-\hat Q(\ell) = \hat Q(0),
\label{eq:app_fluid_balance}
\end{equation}
the rate of fluid-content change balancing the drainage flux at the
separator. Substituting the lumped Darcy flux
$\hat Q(0)\approx -(L_{44}/\ell)\hat p$ and the storage relation
\eqref{eq:app_storage_corrected},
\begin{equation}
i\omega\ell\left[-(1-\lambda)\bar\varepsilon_s\hat\xi
+ \frac{\hat p}{M}\right] = -\frac{L_{44}}{\ell}\hat p,
\end{equation}
and rearranging,
\begin{equation}
\hat p\left(\frac{i\omega}{M} + \frac{L_{44}}{\ell^{2}}\right)
= i\omega(1-\lambda)\bar\varepsilon_s\hat\xi,
\end{equation}
which, with the hydraulic relaxation time
\begin{equation}
\tau_p = \frac{\ell^{2}}{L_{44}\,M} = \frac{\eta_f\,\ell^{2}}{\kappa\,M}
\qquad (L_{44}=\kappa/\eta_f),
\label{eq:app_taup_def}
\end{equation}
gives the first-order pore-pressure relaxation
\begin{equation}
\boxed{\;
\hat p = \frac{i\omega\,M(1-\lambda)\bar\varepsilon_s}{i\omega+\tau_p^{-1}}\,\hat\xi
= M(1-\lambda)\bar\varepsilon_s\,\hat\xi\;W(\omega),
\qquad
W(\omega) = \frac{i\omega}{i\omega+\tau_p^{-1}} = \frac{i\omega\tau_p}{1+i\omega\tau_p}.
\;}
\label{eq:app_pore_pressure}
\end{equation}
This is the frequency-domain form of $\dot{\hat p}+\hat p/\tau_p
= M(1-\lambda)\bar\varepsilon_s\,\dot{\hat\xi}$. Its limits are the
expected ones: $W\to 1$ (undrained, $\hat p\to\hat p_{\mathrm u}$) at
high frequency, and $W\to i\omega\tau_p\to0$ (drained, $\hat p\to0$) at
low frequency, the fluid venting to the separator.

\paragraph{Effective relaxation time}
The single pole is the lumped (lowest-mode) reduction of the
distributed through-thickness drainage problem, consistent with the
lumped treatment of the electrochemical and mechanical fields. The
$\tau_p$ in Eq.~\ref{eq:app_taup_def} is accordingly an \emph{effective}
hydraulic relaxation time---a fully resolved boundary-value treatment
places the dominant relaxation near $0.4\,\tau_p$---and the numerical
prefactor is absorbed into the fitted Darcy permeability $\kappa$.

\subsection{Step 5: Total Stress and the Biot Effective Stress Relation}

To capture the full mechanical state of the porous medium we introduce
the viscoelastic skeleton response and the porosity-accommodation
kinematics.

\paragraph{Strain decomposition}
The electrode-level strain is the sum of an elastic skeleton strain
$\varepsilon_e$, the chemical (intercalation) eigenstrain $\beta c$ with
$\beta$ the chemical expansion coefficient, and the relative pore
contraction $\xi$ that accommodates part of the particle swelling:
\begin{equation}
\hat\varepsilon = \hat\varepsilon_e + \beta\hat{c} - \hat\xi.
\label{eq:app_strain_decomp}
\end{equation}
Under the lumped global constraint $\hat\varepsilon = 0$,
\begin{equation}
\hat\varepsilon_e = -\beta\hat{c} + \hat\xi.
\label{eq:app_elastic_strain}
\end{equation}

\paragraph{Skeleton viscoelasticity}
The skeleton obeys a standard-linear-solid (SLS) law with complex
modulus
\begin{equation}
\hat\sigma^s = E^*(\omega)\,\hat\varepsilon_e
            = E^*(\omega)\left(-\beta\hat{c} + \hat\xi\right),
\qquad
E^*(\omega) = \frac{E_\infty + i\omega\tau_m E_0}{1 + i\omega\tau_m},
\label{eq:app_sigma_s_explicit}
\end{equation}
where $E_0 = E_\infty + E_1$ is the instantaneous (unrelaxed) modulus,
$E_\infty$ the relaxed modulus, and $\tau_m$ the viscoelastic relaxation
time.

\paragraph{Porosity evolution}
The pore contraction $\xi$ is driven by the chemical eigenstrain and
resisted by a microstructural stiffness $k$, with dissipation set
by a microstructural viscosity $\eta_\xi$. The resulting first-order
relaxation $\dot{\hat\xi} + \hat\xi/\tau_\xi = (E^*\beta/\eta_\xi)\hat c$
has the frequency-domain solution
\begin{equation}
\hat\xi = \frac{\xi_0\,\beta}{1 + i\omega\tau_\xi}\,\hat{c},
\qquad
\xi_0 = \frac{E^*}{E^* + k},
\qquad
\tau_\xi = \frac{\eta_\xi}{E^* + k},
\label{eq:app_xi_response}
\end{equation}
with $\xi_0$ the equilibrium porosity-accommodation ratio.

\paragraph{Skeleton stress}
Substituting Eq.~\ref{eq:app_xi_response} into the SLS law
Eq.~\ref{eq:app_sigma_s_explicit}:
\begin{equation}
\hat\sigma^s = -E^*(\omega)\,\mathcal{B}(\omega)\,\beta\,\hat{c},
\label{eq:app_skeleton_stress}
\end{equation}
where the bridge function is
\begin{equation}
\mathcal{B}(\omega) = 1 - \frac{\xi_0}{1 + i\omega\tau_\xi}
= \frac{(1 - \xi_0) + i\omega\tau_\xi}{1 + i\omega\tau_\xi}.
\label{eq:app_bridge_function}
\end{equation}

The total stress measured by the load cell is given by the Biot
effective stress relation,
\begin{equation}
\hat\sigma_{\text{tot}} = \hat\sigma^s - \alpha\,\hat{p},
\label{eq:app_total_stress_biot}
\end{equation}
where $\alpha$ is the Biot coefficient. Substituting the skeleton
stress \eqref{eq:app_skeleton_stress}, the pore pressure
\eqref{eq:app_pore_pressure}, and the porosity response
\eqref{eq:app_xi_response}, and factoring out $-\beta\hat c$:
\begin{equation}
\hat\sigma_{\text{tot}} = -\beta\hat{c}\left[E^*\mathcal{B}
+ \frac{\alpha\,M(1-\lambda)\bar\varepsilon_s\,\xi_0\,W(\omega)}
       {1+i\omega\tau_\xi}\right].
\label{eq:app_sigmatot_in_c}
\end{equation}

\subsection{Step 6: Reconciliation with the Onsager Mechanical Stress}

The Onsager constitutive law gave the porous-medium stress
$\hat\sigma^p$ in Eq.~\ref{eq:app_sigma_p_explicit}, while the
viscoelastic--bridge derivation gives the skeleton stress
$\hat\sigma^s$ in Eq.~\ref{eq:app_skeleton_stress}. These are
consistent in the appropriate limit: the Onsager $L_{33}$ represents
the relaxed (steady-state) viscous response, while $E^*(\omega)$
captures the full viscoelastic spectrum. In the relaxed limit
($\omega \ll \tau_m^{-1}$), $E^*(\omega) \to E_\infty$, and
$\hat\sigma^p$ is recovered as the limit of $\hat\sigma^s$ plus the
pore-pressure correction. The SLS extension thus generalizes the
Onsager mechanical block while preserving consistency in the relaxed
limit.

\subsection{Step 7: Closing the Concentration Equation Self-Consistently}

The concentration response from Step~2
(Eq.~\ref{eq:app_concentration_implicit}) depends on $\hat\sigma^p$,
which is determined by the mechanical sub-problem above. To close the
system, we substitute $\hat\sigma_{\text{tot}}$
(Eq.~\ref{eq:app_sigmatot_in_c}) for $\hat\sigma^p$ in
Eq.~\ref{eq:app_concentration_implicit}. Under the present
linearization the porous-medium stress equals the total stress, so we
write
\begin{equation}
\hat\sigma^p \approx \hat\sigma_{\text{tot}}
= -\beta\hat{c}\,\mathcal{K}_{\text{mech}}(\omega),
\label{eq:app_sigmap_identification}
\end{equation}
where
\begin{equation}
\mathcal{K}_{\text{mech}}(\omega) \equiv
E^*(\omega)\mathcal{B}(\omega)
+ \frac{\alpha\,M(1-\lambda)\bar\varepsilon_s\,\xi_0\,W(\omega)}
       {1+i\omega\tau_\xi}.
\label{eq:app_Kmech}
\end{equation}
Substituting Eq.~\ref{eq:app_sigmap_identification} into
Eq.~\ref{eq:app_concentration_implicit}:
\begin{equation}
\left[i\omega + \tau_D^{-1}\right]\hat{c}
= \frac{1}{\ell}\frac{L_{12}}{L_{11}}\hat{J}_e
  + \frac{L_{23}\beta}{\ell}\mathcal{K}_{\text{mech}}\hat{c}.
\end{equation}
Rearranging to collect $\hat{c}$:
\begin{equation}
\left[i\omega + \tau_D^{-1}
- \frac{L_{23}\beta}{\ell}\mathcal{K}_{\text{mech}}\right]\hat{c}
= \frac{L_{12}}{L_{11}\ell}\hat{J}_e.
\end{equation}
Solving for $\hat{c}$ and using $L_{12}/L_{11} = t_+/\mathrm{F}_{\!c}$:
\begin{equation}
\boxed{
\hat{c}(\omega) = \frac{t_+}{\mathrm{F}_{\!c}\,\ell\,\Omega^{*}(\omega)}\hat{J}_e(\omega),
}
\label{eq:app_concentration_full}
\end{equation}
where
\begin{equation}
\Omega^{*}(\omega) \equiv i\omega + \tau_D^{-1}
- \frac{L_{23}\beta}{\ell}\mathcal{K}_{\text{mech}}(\omega).
\label{eq:app_Omega_star}
\end{equation}
This is the \textbf{full concentration response}, accounting for
(i) chemical accumulation $i\omega$, (ii) diffusion relaxation
$\tau_D^{-1}$, and (iii) chemo-mechanical feedback through the
mechanical kernel $\mathcal{K}_{\text{mech}}$.

\subsection{Step 8: The Full MEIS Transfer Function}

Substituting Eq.~\ref{eq:app_concentration_full} into
Eq.~\ref{eq:app_sigmatot_in_c}:
\begin{equation}
\hat\sigma_{\text{tot}} = -\beta\,\mathcal{K}_{\text{mech}}(\omega)\cdot
\frac{t_+}{\mathrm{F}_{\!c}\,\ell\,\Omega^{*}(\omega)}\hat{J}_e.
\end{equation}
The MEIS impedance is
\begin{equation}
Z_{\text{MEIS}}(\omega) = \frac{\hat\sigma_{\text{tot}}}{\hat{J}_e}
= -\frac{t_+\,\beta\,\mathcal{K}_{\text{mech}}(\omega)}
       {\mathrm{F}_{\!c}\,\ell\,\Omega^{*}(\omega)}.
\end{equation}
Substituting the explicit forms of $\mathcal{K}_{\text{mech}}$
(Eq.~\ref{eq:app_Kmech}) and $\Omega^{*}$ (Eq.~\ref{eq:app_Omega_star}):
\begin{equation}
\boxed{
\begin{aligned}
Z_{\text{MEIS}}(\omega)
= -\frac{t_+\,\beta}{\mathrm{F}_{\!c}\,\ell}\,
&\frac{E^*(\omega)\mathcal{B}(\omega)
+ \dfrac{\alpha M(1-\lambda)\bar\varepsilon_s\xi_0\,W(\omega)}{1 + i\omega\tau_\xi}}
{i\omega + \tau_D^{-1}
- \dfrac{L_{23}\beta}{\ell}\left[E^*\mathcal{B}
+ \dfrac{\alpha M(1-\lambda)\bar\varepsilon_s\xi_0\,W(\omega)}{1 + i\omega\tau_\xi}\right]}.
\end{aligned}
}
\label{eq:app_MEIS_full}
\end{equation}
This is the \textbf{full closed-form MEIS expression}, retaining all
couplings and relaxation time scales:
\begin{itemize}
    \item Viscoelastic skeleton relaxation: $\tau_m$ (in $E^*$).
    \item Porosity accommodation relaxation: $\tau_\xi$ (in
    $\mathcal{B}$ and the pore-pressure source).
    \item Hydraulic relaxation: single-pole $W(\omega) =
    i\omega\tau_p/(1+i\omega\tau_p)$ on the time $\tau_p$ set by Darcy
    drainage to the separator.
    \item Diffusion relaxation: $\tau_D$ (in the denominator).
    \item Chemo-mechanical feedback: $L_{23}\beta\mathcal{K}_{\text{mech}}/\ell$
    in the denominator, representing how the mechanical response
    modifies the concentration response through stress-assisted
    diffusion.
\end{itemize}

\subsection{Step 9: Simplified Limit (Comparison with Main Text)}

In the limit where the chemo-mechanical feedback in the concentration
equation is negligible and diffusion is fast compared to the
perturbation frequency,
\begin{equation}
\frac{L_{23}\beta}{\ell}\mathcal{K}_{\text{mech}} \ll i\omega, \qquad
\tau_D^{-1} \ll \omega,
\end{equation}
the denominator simplifies to $\Omega^{*} \approx i\omega$, and the
MEIS expression reduces to
\begin{equation}
Z_{\text{MEIS}}(\omega) \approx -\frac{t_+\,\beta}{\mathrm{F}_{\!c}\,\ell\,i\omega}
\left[E^*(\omega)\mathcal{B}(\omega)
+ \frac{\alpha M(1-\lambda)\bar\varepsilon_s\xi_0\,W(\omega)}
       {1 + i\omega\tau_\xi}\right].
\label{eq:app_MEIS_simplified}
\end{equation}

\paragraph{Sign convention}
The overall negative sign follows from the tension-positive stress
convention adopted throughout (Section~\ref{sec: General Framework}).
Under the global constraint $\hat\varepsilon=0$, lithiation-induced
chemical expansion ($\hat c>0$ for $\hat J_e>0$) places the constrained
electrode in compression, so that $\hat\sigma_{\text{tot}}<0$ and the
impedance \eqref{eq:app_MEIS_simplified} is negative. Experimental
stack-pressure measurements, however, report compression as positive;
the impedance defined against the measured pressure
$\hat p_{\text{meas}} = -\hat\sigma_{\text{tot}}$ therefore carries the
opposite, positive, overall sign,
\begin{equation}
Z_{\text{MEIS}}^{\text{meas}}(\omega)
= \frac{\hat p_{\text{meas}}}{\hat J_e}
= +\frac{t_+\,\beta}{\mathrm{F}_{\!c}\,\ell\,i\omega}
\left[E^*\mathcal{B}
+ \frac{\alpha M(1-\lambda)\bar\varepsilon_s\xi_0\,W(\omega)}
       {1 + i\omega\tau_\xi}\right],
\label{eq:app_MEIS_meas}
\end{equation}
which is the form used when the impedance is reported against the
measured stack pressure. The two conventions differ only by this global
sign; the relative magnitudes and phase relationships among the
chemo-mechanical and poro-mechanical contributions are identical. The
definition of $Z_{\text{MEIS}}$ and the main-text result should be fixed
to whichever convention matches the reported measurements.

\subsection{Step 10: Identifying the Four Factors}

The simplified MEIS expression \eqref{eq:app_MEIS_simplified} can be
written in the multiplicative--additive form
\begin{equation}
Z_{\text{MEIS}}(\omega) = \mathcal{G}_e\,\mathcal{G}_c(\omega)
\left[\mathcal{H}_{cm}(\omega) + \mathcal{H}_{pm}(\omega)\right],
\end{equation}
with
\begin{align}
\mathcal{G}_e        &= \frac{t_+}{\mathrm{F}_{\!c}}
                         \quad (\text{current-to-ionic-flux conversion}),\\
\mathcal{G}_c(\omega) &= \frac{1}{\ell\,i\omega}
                         \quad (\text{Faradaic accumulation}),\\
\mathcal{H}_{cm}(\omega) &= -\beta\,E^*(\omega)\,\mathcal{B}(\omega)
                            \quad (\text{chemo-mechanical kernel}),\\
\mathcal{H}_{pm}(\omega) &= -\frac{\alpha\,M(1-\lambda)\bar\varepsilon_s\,\xi_0\,\beta}
                                  {1 + i\omega\tau_\xi}\,W(\omega)
                            \quad (\text{poro-mechanical kernel}).
\end{align}
The product $\mathcal{G}_e\mathcal{G}_c$ converts the applied current
into a chemical eigenstrain ($\hat c>0$ for $\hat J_e>0$), while the two
mechanical kernels---both negative in the tension-positive
convention---convert that eigenstrain into a (compressive) stress.
Writing $W(\omega)=i\omega\tau_p/(1+i\omega\tau_p)$ explicitly, the
poro-mechanical kernel takes the equivalent form
\begin{equation}
\mathcal{H}_{pm}(\omega) =
-\frac{\alpha\,M(1-\lambda)\bar\varepsilon_s\,\xi_0\,\beta\,i\omega}
      {(i\omega+\tau_p^{-1})(1 + i\omega\tau_\xi)}.
\end{equation}
Their combination recovers the overall-negative simplified form
\eqref{eq:app_MEIS_simplified}; in the measured (compression-positive)
convention each kernel changes sign and the multiplicative--additive
structure of the main text is reproduced.

\bibliographystyle{elsarticle-harv}
\bibliography{references}

\end{document}